\documentclass[longauth]{aa}  
\usepackage[dvipsnames,svgnames]{xcolor}
\usepackage{blindtext}
\usepackage{xspace}
\usepackage{amssymb}
\usepackage{amsmath}
\usepackage[T1]{fontenc}
\usepackage{mathtools}
\DeclareRobustCommand{\VAN}[3]{#2}
\let\VANthebibliography\thebibliography
\def\thebibliography{\DeclareRobustCommand{\VAN}[3]{##3}\VANthebibliography}
\usepackage{gensymb}
\usepackage{wasysym}
\usepackage{graphicx}
\usepackage[export]{adjustbox}
\usepackage{verbatim} 
\usepackage{multirow}
\usepackage{caption} 
\usepackage{soul}
\usepackage[colorlinks=true,linkcolor=blue]{hyperref}
\hypersetup{citecolor=blue, urlcolor=blue}
\usepackage{caption}
\usepackage{subcaption}
\usepackage{float}
\usepackage[commandnameprefix=always,commentmarkup=footnote]{changes}
\usepackage{ulem}
\usepackage{physics}
\usepackage[toc,page]{appendix}
\usepackage{pdflscape}
\usepackage[export]{adjustbox}
\usepackage{afterpage}

\DeclareUnicodeCharacter{2212}{-}

\def\flat{\textit{Fermi}-LAT\xspace}

\def\swift{\textit{Swift}\xspace}

\def\hess{H.E.S.S.\xspace}
\def\psrb{PSR~B1259-63/LS~2883\xspace}

\def\hessJ{HESS J1303–631\xspace}
\def\tp{\ensuremath{t_{\mathrm{p}}}}
\def\ergs{\ensuremath{\mathrm{erg}\,\mathrm{s}^{-1}}}
\usepackage[switch]{lineno}

\begin{document}

   \title{\hess{} observations of the 2021 periastron passage of \psrb}

\author{H.E.S.S. Collaboration
\and F.~Aharonian \inst{\ref{DIAS},\ref{MPIK},\ref{Yerevan}}
\and F.~Ait~Benkhali \inst{\ref{LSW}}
\and J.~Aschersleben \inst{\ref{Groningen}}
\and H.~Ashkar \inst{\ref{LLR}}
\and M.~Backes \inst{\ref{UNAM},\ref{NWU}}
\and V.~Barbosa~Martins \inst{\ref{DESY}}
\and R.~Batzofin \inst{\ref{UP}}
\and Y.~Becherini \inst{\ref{APC},\ref{Linnaeus}}
\and D.~Berge \inst{\ref{DESY},\ref{HUB}}
\and K.~Bernl\"ohr \inst{\ref{MPIK}}
\and M.~B\"ottcher \inst{\ref{NWU}}
\and C.~Boisson \inst{\ref{LUTH}}
\and J.~Bolmont \inst{\ref{LPNHE}}
\and M.~de~Bony~de~Lavergne \inst{\ref{CEA}}
\and J.~Borowska \inst{\ref{HUB}}
\and M.~Bouyahiaoui \inst{\ref{MPIK}}
\and R.~Brose \inst{\ref{DIAS}}
\and A.~Brown \inst{\ref{Oxford}}
\and F.~Brun \inst{\ref{CEA}}
\and B.~Bruno \inst{\ref{ECAP}}
\and T.~Bulik \inst{\ref{UWarsaw}}
\and C.~Burger-Scheidlin \inst{\ref{DIAS}}
\and S.~Caroff \inst{\ref{LAPP}}
\and S.~Casanova \inst{\ref{IFJPAN}}
\and J.~Celic \inst{\ref{ECAP}}
\and M.~Cerruti \inst{\ref{APC}}
\and T.~Chand \inst{\ref{NWU}}
\and S.~Chandra \inst{\ref{NWU}}
\and A.~Chen \inst{\ref{Wits}}
\and J.~Chibueze \inst{\ref{NWU}}
\and O.~Chibueze \inst{\ref{NWU}}
\and G.~Cotter \inst{\ref{Oxford}}
\and J.~Damascene~Mbarubucyeye \inst{\ref{DESY}}
\and J.~Devin \inst{\ref{LUPM}}
\and J.~Djuvsland \inst{\ref{MPIK}}
\and A.~Dmytriiev \inst{\ref{NWU}}
\and K.~Egberts \inst{\ref{UP}}
\and S.~Einecke \inst{\ref{Adelaide}}
\and J.-P.~Ernenwein \inst{\ref{CPPM}}
\and G.~Fontaine \inst{\ref{LLR}}
\and S.~Funk \inst{\ref{ECAP}}
\and S.~Gabici \inst{\ref{APC}}
\and Y.A.~Gallant \inst{\ref{LUPM}}
\and D.~Glawion \inst{\ref{ECAP}}
\and J.F.~Glicenstein \inst{\ref{CEA}}
\and P.~Goswami \inst{\ref{APC}}
\and G.~Grolleron \inst{\ref{LPNHE}}
\and L.~Haerer \inst{\ref{MPIK}}
\and B.~He\ss \inst{\ref{IAAT}}
\and W.~Hofmann \inst{\ref{MPIK}}
\and T.~L.~Holch \inst{\ref{DESY}}
\and M.~Holler \inst{\ref{Innsbruck}}
\and Zhiqiu~Huang \inst{\ref{MPIK}}
\and M.~Jamrozy \inst{\ref{UJK}}
\and F.~Jankowsky \inst{\ref{LSW}}
\and V.~Joshi \inst{\ref{ECAP}}
\and I.~Jung-Richardt \inst{\ref{ECAP}}
\and E.~Kasai \inst{\ref{UNAM}}
\and K.~Katarzy{\'n}ski \inst{\ref{NCUT}}
\and D.~Khangulyan \inst{\ref{Rikkyo}}
\and R.~Khatoon \inst{\ref{NWU}}
\and B.~Kh\'elifi \inst{\ref{APC}}
\and W.~Klu\'{z}niak \inst{\ref{NCAC}}
\and Nu.~Komin \inst{\ref{Wits}}
\and K.~Kosack \inst{\ref{CEA}}
\and D.~Kostunin \inst{\ref{DESY}}
\and A.~Kundu \inst{\ref{NWU}}
\and R.G.~Lang \inst{\ref{ECAP}}
\and S.~Le~Stum \inst{\ref{CPPM}}
\and F.~Leitl \inst{\ref{ECAP}}
\and A.~Lemi\`ere \inst{\ref{APC}}
\and M.~Lemoine-Goumard \inst{\ref{CENBG}}
\and J.-P.~Lenain \inst{\ref{LPNHE}}
\and F.~Leuschner \inst{\ref{IAAT}}
\and J.~Mackey \inst{\ref{DIAS}}
\and D.~Malyshev \inst{\ref{IAAT}}$^{,}$\footnotemark[1]
\and G.~Mart\'i-Devesa \inst{\ref{Innsbruck}}
\and R.~Marx \inst{\ref{LSW}}
\and A.~Mehta \inst{\ref{DESY}}
\and P.J.~Meintjes \inst{\ref{UFS}}
\and A.~Mitchell \inst{\ref{ECAP}}
\and R.~Moderski \inst{\ref{NCAC}}
\and L.~Mohrmann \inst{\ref{MPIK}}
\and A.~Montanari \inst{\ref{LSW}}
\and E.~Moulin \inst{\ref{CEA}}
\and T.~Murach \inst{\ref{DESY}}
\and M.~de~Naurois \inst{\ref{LLR}}
\and J.~Niemiec \inst{\ref{IFJPAN}}
\and S.~Ohm \inst{\ref{DESY}}
\and E.~de~Ona~Wilhelmi \inst{\ref{DESY}}
\and M.~Ostrowski \inst{\ref{UJK}}
\and S.~Panny \inst{\ref{Innsbruck}}
\and M.~Panter \inst{\ref{MPIK}}
\and R.D.~Parsons \inst{\ref{HUB}}
\and U.~Pensec \inst{\ref{LPNHE}}
\and G.~Peron \inst{\ref{APC}}
\and D.A.~Prokhorov \inst{\ref{Amsterdam}}
\and G.~P\"uhlhofer \inst{\ref{IAAT}}$^{,}$\footnotemark[1]
\and M.~Punch \inst{\ref{APC}}
\and A.~Quirrenbach \inst{\ref{LSW}}
\and M.~Regeard \inst{\ref{APC}}
\and A.~Reimer \inst{\ref{Innsbruck}}
\and O.~Reimer \inst{\ref{Innsbruck}}
\and I.~Reis \inst{\ref{CEA}}
\and H.~Ren \inst{\ref{MPIK}}
\and F.~Rieger \inst{\ref{MPIK}}
\and B.~Rudak \inst{\ref{NCAC}}
\and E.~Ruiz-Velasco \inst{\ref{MPIK}}
\and V.~Sahakian \inst{\ref{Yerevan}}
\and H.~Salzmann \inst{\ref{IAAT}}
\and A.~Santangelo \inst{\ref{IAAT}}
\and M.~Sasaki \inst{\ref{ECAP}}
\and J.~Sch\"afer \inst{\ref{ECAP}}
\and F.~Sch\"ussler \inst{\ref{CEA}}
\and H.M.~Schutte \inst{\ref{NWU}}
\and J.N.S.~Shapopi \inst{\ref{UNAM}}
\and S.~Spencer \inst{\ref{ECAP}}
\and {\L.}~Stawarz \inst{\ref{UJK}}
\and R.~Steenkamp \inst{\ref{UNAM}}
\and S.~Steinmassl \inst{\ref{MPIK}}
\and C.~Steppa \inst{\ref{UP}}
\and K.~Streil \inst{\ref{ECAP}}
\and I.~Sushch \inst{\ref{NWU}}$^{,}$\footnotemark[1]
\and T.~Takahashi \inst{\ref{KAVLI}}
\and T.~Tanaka \inst{\ref{Konan}}
\and A.M.~Taylor \inst{\ref{DESY}}
\and R.~Terrier \inst{\ref{APC}}
\and C.~Thorpe-Morgan \inst{\ref{IAAT}}$^{,}$\thanks{Corresponding authors;\newline\email{\href{mailto:contact.hess@hess-experiment.eu}{contact.hess@hess-experiment.eu}}}
\and M.~Tluczykont \inst{\ref{UHH}}
\and T.~Unbehaun \inst{\ref{ECAP}}
\and C.~van~Eldik \inst{\ref{ECAP}}
\and B.~van~Soelen \inst{\ref{UFS}}
\and M.~Vecchi \inst{\ref{Groningen}}
\and C.~Venter \inst{\ref{NWU}}
\and J.~Vink \inst{\ref{Amsterdam}}
\and T.~Wach \inst{\ref{ECAP}}
\and S.J.~Wagner \inst{\ref{LSW}}
\and F.~Werner \inst{\ref{MPIK}}
\and A.~Wierzcholska \inst{\ref{IFJPAN}}
\and M.~Zacharias \inst{\ref{LSW},\ref{NWU}}
\and A.A.~Zdziarski \inst{\ref{NCAC}}
\and A.~Zech \inst{\ref{LUTH}}
\and N.~\.Zywucka \inst{\ref{NWU}}
}

\institute{
Dublin Institute for Advanced Studies, 31 Fitzwilliam Place, Dublin 2, Ireland \label{DIAS} \and
Max-Planck-Institut f\"ur Kernphysik, P.O. Box 103980, D 69029 Heidelberg, Germany \label{MPIK} \and
Yerevan State University,  1 Alek Manukyan St, Yerevan 0025, Armenia \label{Yerevan} \and
Landessternwarte, Universit\"at Heidelberg, K\"onigstuhl, D 69117 Heidelberg, Germany \label{LSW} \and
Kapteyn Astronomical Institute, University of Groningen, Landleven 12, 9747 AD Groningen, The Netherlands \label{Groningen} \and
Laboratoire Leprince-Ringuet, École Polytechnique, CNRS, Institut Polytechnique de Paris, F-91128 Palaiseau, France \label{LLR} \and
University of Namibia, Department of Physics, Private Bag 13301, Windhoek 10005, Namibia \label{UNAM} \and
Centre for Space Research, North-West University, Potchefstroom 2520, South Africa \label{NWU} \and
Deutsches Elektronen-Synchrotron DESY, Platanenallee 6, 15738 Zeuthen, Germany \label{DESY} \and
Institut f\"ur Physik und Astronomie, Universit\"at Potsdam,  Karl-Liebknecht-Strasse 24/25, D 14476 Potsdam, Germany \label{UP} \and
Université de Paris, CNRS, Astroparticule et Cosmologie, F-75013 Paris, France \label{APC} \and
Department of Physics and Electrical Engineering, Linnaeus University,  351 95 V\"axj\"o, Sweden \label{Linnaeus} \and
Institut f\"ur Physik, Humboldt-Universit\"at zu Berlin, Newtonstr. 15, D 12489 Berlin, Germany \label{HUB} \and
Laboratoire Univers et Théories, Observatoire de Paris, Université PSL, CNRS, Université Paris Cité, 5 Pl. Jules Janssen, 92190 Meudon, France \label{LUTH} \and
Sorbonne Universit\'e, CNRS/IN2P3, Laboratoire de Physique Nucl\'eaire et de Hautes Energies, LPNHE, 4 place Jussieu, 75005 Paris, France \label{LPNHE} \and
IRFU, CEA, Universit\'e Paris-Saclay, F-91191 Gif-sur-Yvette, France \label{CEA} \and
University of Oxford, Department of Physics, Denys Wilkinson Building, Keble Road, Oxford OX1 3RH, UK \label{Oxford} \and
Friedrich-Alexander-Universit\"at Erlangen-N\"urnberg, Erlangen Centre for Astroparticle Physics, Nikolaus-Fiebiger-Str. 2, 91058 Erlangen, Germany \label{ECAP} \and
Astronomical Observatory, The University of Warsaw, Al. Ujazdowskie 4, 00-478 Warsaw, Poland \label{UWarsaw} \and
Université Savoie Mont Blanc, CNRS, Laboratoire d'Annecy de Physique des Particules - IN2P3, 74000 Annecy, France \label{LAPP} \and
Instytut Fizyki J\c{a}drowej PAN, ul. Radzikowskiego 152, 31-342 Krak{\'o}w, Poland \label{IFJPAN} \and
School of Physics, University of the Witwatersrand, 1 Jan Smuts Avenue, Braamfontein, Johannesburg, 2050 South Africa \label{Wits} \and
Laboratoire Univers et Particules de Montpellier, Universit\'e Montpellier, CNRS/IN2P3,  CC 72, Place Eug\`ene Bataillon, F-34095 Montpellier Cedex 5, France \label{LUPM} \and
School of Physical Sciences, University of Adelaide, Adelaide 5005, Australia \label{Adelaide} \and
Aix Marseille Universit\'e, CNRS/IN2P3, CPPM, Marseille, France \label{CPPM} \and
Institut f\"ur Astronomie und Astrophysik, Universit\"at T\"ubingen, Sand 1, D 72076 T\"ubingen, Germany \label{IAAT} \and
Universit\"at Innsbruck, Institut f\"ur Astro- und Teilchenphysik, Technikerstraße 25, 6020 Innsbruck, Austria \label{Innsbruck} \and
Obserwatorium Astronomiczne, Uniwersytet Jagiello{\'n}ski, ul. Orla 171, 30-244 Krak{\'o}w, Poland \label{UJK} \and
Institute of Astronomy, Faculty of Physics, Astronomy and Informatics, Nicolaus Copernicus University,  Grudziadzka 5, 87-100 Torun, Poland \label{NCUT} \and
Department of Physics, Rikkyo University, 3-34-1 Nishi-Ikebukuro, Toshima-ku, Tokyo 171-8501, Japan \label{Rikkyo} \and
Nicolaus Copernicus Astronomical Center, Polish Academy of Sciences, ul. Bartycka 18, 00-716 Warsaw, Poland \label{NCAC} \and
Universit\'e Bordeaux, CNRS, LP2I Bordeaux, UMR 5797, F-33170 Gradignan, France \label{CENBG} \and
Department of Physics, University of the Free State,  PO Box 339, Bloemfontein 9300, South Africa \label{UFS} \and
GRAPPA, Anton Pannekoek Institute for Astronomy, University of Amsterdam,  Science Park 904, 1098 XH Amsterdam, The Netherlands \label{Amsterdam} \and
Kavli Institute for the Physics and Mathematics of the Universe (WPI), The University of Tokyo Institutes for Advanced Study (UTIAS), The University of Tokyo, 5-1-5 Kashiwa-no-Ha, Kashiwa, Chiba, 277-8583, Japan \label{KAVLI} \and
Department of Physics, Konan University, 8-9-1 Okamoto, Higashinada, Kobe, Hyogo 658-8501, Japan \label{Konan} \and
Universit\"at Hamburg, Institut f\"ur Experimentalphysik, Luruper Chaussee 149, D 22761 Hamburg, Germany \label{UHH}
}

\abstract{\psrb is a gamma-ray binary system that hosts a pulsar in an eccentric orbit, with a 3.4 year period, around an O9.5Ve star (LS~2883). At orbital phases close to periastron passages, the system radiates bright and variable non-thermal emission, for which the temporal and spectral properties of this emission are, for now, poorly understood. In this regard, very high-energy (VHE) emission is especially useful to study and constrain radiation processes and particle acceleration in the system. We report on an extensive VHE observation campaign conducted with the High Energy Stereoscopic System, comprised of approximately 100 hours of data taken over five months, from $t_{\mathrm{p}}-24$~days to $t_{\mathrm{p}}+127$~days around the system's 2021 periastron passage (where $t_{\mathrm{p}}$ is the time of periastron). We also present the timing and spectral analyses of the source. The VHE light curve in 2021 is consistent overall with the stacked light curve of all previous observations. Within the light curve, we report a VHE maximum at times coincident with the third X-ray peak first detected in the 2021 X-ray light curve. In the light curve -- although sparsely sampled in this time period -- we see no VHE enhancement during the second disc crossing. In addition, we see no correspondence to the 2021 GeV flare in the VHE light curve. The VHE spectrum obtained from the analysis of the 2021 dataset is best described by a power law of spectral index $\Gamma = 2.65 ~\pm~ 0.04_{\text{stat}}$ $~\pm~0.04_{\text{sys}}$, a value consistent with the spectral index obtained from the analysis of data collected with H.E.S.S. during the previous observations of the source. We report spectral variability with a difference of $\Delta \Gamma = 0.56 ~\pm~ 0.18_{\text{stat}}$ $~\pm~0.10_{\text{sys}}$ at $95 \%$ confidence intervals, between sub-periods of the 2021 dataset. We also detail our investigation into  X-ray/TeV and GeV/TeV flux correlations in the 2021 periastron passage. We find a linear correlation between contemporaneous flux values of X-ray and TeV datasets, detected mainly after $t_{\mathrm{p}}+25$~days, suggesting a change in the available energy for non-thermal radiation processes. We detect no significant correlation between GeV and TeV flux points, within the uncertainties of the measurements, from  $\sim t_p-23$ ~days to  $\sim t_p+126$ ~days. This suggests that the GeV and TeV emission originate from different electron populations.}

\maketitle
\section{Introduction}
\label{Section:Introduction}

Gamma-ray loud binaries (GRLBs) are a subclass of high-mass and intermediate-mass binary systems characterised by their energy spectra peaking above $1$~MeV, but typically at $E\gtrsim100$~MeV, and extending to beyond~$1$ TeV. While hundreds of high-mass binaries have been detected in the X-ray band, the current generation of Cherenkov telescopes and gamma-ray satellites have only been able to detect about a dozen GRLB systems ~\citep[see, e.g.,][for recent reviews]{dubus13,grlbcta19}. The physical environments and mechanisms leading to the production of such energetic radiation in these systems are not firmly established.

GRLB systems are comprised of a massive early-type star (spectral class O or B) and a compact object (a neutron star or a black hole). The nature of this compact object is difficult to discern in the majority of cases, in several systems however, the compact object has been identified as a non-accreting pulsar such as in the cases of \psrb, PSR~J2032+4127 and LS~I+61 303~\citep{johnston92, abdo_2009_J2032_detection, weng22}. Additionally, evidence of hard X-ray pulsations have been reported in the system LS~5039~\citep{yoneda_2020}, tentatively suggesting a neutron star companion as well.

The \psrb system was discovered during a high-frequency radio survey intending to search for nearby pulsars~\citep{johnston92}. Subsequent radio and optical observations resulted in the identification of the compact object in the system as a young radio pulsar (spin period $\sim 48$~ms), in a highly eccentric $(e=0.87)$ 3.4-year ($1236.724526 \pm 6 \times 10^{-6}$ day) orbit around the O9.5Ve star LS~2883~\citep{johnston92,johnston94,Negueruela_2011,shannon14, millerjones_2018}.\footnote{In the following, we assume that the 2021 periastron of \psrb occurred at $t_{\mathrm{p}}=59254.867359$ MJD.} The system is located at a distance of $2.39 ~\pm~ 0.19$ kpc from Earth~\citep{gaiacollab_2018}, and recent measurements of the inclination angle suggest that the binary orbit is observed at an angle of $154^\circ$ to the line of sight ~\citep{millerjones_2018}. The projected semi-major axis is $a\sin i=1296.27448\pm0.00014$~lt-s~\citep{millerjones_2018}, which for the pulsar's orbital eccentricity corresponds to apastron and periastron separations of $11$ AU and $0.8$ AU, respectively. Additionally, the spin-down luminosity of the pulsar was estimated to be $L_{\mathrm{sd}} = 8.2\times10^{35}$~erg\,s$^{-1}$~\citep{johnston94}, with a characteristic age of $330$ kyr~\citep{johnston2005}.

The companion star LS~2883 has a bolometric luminosity of $L_{*} = 2.3\times10^{38}$~erg\,s$^{-1}$~\citep{Negueruela_2011} and hosts a decretion disc that extends up to at least 20 stellar radii~\citep{johnston92,Negueruela_2011,chernyakova_14} from the star ($0.56$ AU). The radius of LS~2883 is about $10$\(\textup{R}_\odot\) ($0.05$ AU) \citep{Negueruela_2011}, and its mass is $\sim 24$ \(\textup{M}_\odot\)\citep{shannon14}. 
The disappearance of pulsed radio emission at $\sim t_{\mathrm{p}} -16$ days (where $t_{\mathrm{p}}$ is the time of periastron), and its reappearance at $\sim t_{\mathrm{p}} +16$~days~\citep{johnston2005}, as well as observations of the dispersion measure along the periastron passage, both suggest that the stellar decretion disc is inclined with respect to the orbital plane ~\citep{johnson96}. Measurements of this inclination angle between the plane of the pulsar's orbit and the circumstellar disc suggest an angle of $\sim 35^\circ$ \citep{johnston94,shannon14}.   

Following its optical and radio detection, the system was later detected in the X-ray band with the \textit{ROSAT} satellite ~\citep{cominsky94}. In the X-ray regime, \psrb is detected during its entire orbit with a non-thermal, non-pulsed spectrum~\citep{marino_2023}. While the X-ray flux level is minimal around apastron, close to the periastron passage the keV light curve is typically characterised by two maxima roughly coinciding with the times of the disappearance and re-appearance of pulsed radio emission~\citep[see e.g.][]{chernyakova15}. These peaks are usually interpreted as being connected to the pulsar crossing the Oe stellar disc. During the 2021 periastron passage, the X-ray light curve exhibited a third maximum between $\sim t_p+30$ and $t_{\mathrm{p}}+50$~days ~\citep{chernyakova_21} (henceforth referred to as the third X-ray peak), in addition to the two X-ray peaks at $\sim t_p ~\pm~ 16$~days detected in all observed periastron passages.

\psrb was detected in the GeV band with \flat~\citep{Abdo2011_b1259, tam11}. 
At these energies, the system is characterised by a relatively low flux level in the period between $t_{\mathrm{p}}-30$ and $t_{\mathrm{p}}+30$~days. It later enters a high flux state (coined as the ``GeV flare'') that has been detected following all periastron passages observed with \flat to date~\citep[$2010-2021$][]{Abdo2011_b1259, vansoelen_2015, caliandro_2015, wood_2018, chang_2021}. However, for all periastron passages to date during which very high-energy (VHE; $\gtrsim 100$ GeV) observations were taken contemporaneously with the corresponding GeV flare, no clear counterpart at very high energies has been seen~\citep{psrb2011, hess_psrb2020}. The GeV flare in 2017 began after a noticeable delay, starting at up to $\sim 50$~days after the periastron passage ~\citep{chang_2021}. The light curve of the GeV flare obtained from the 2017 periastron passage also showed a number of extremely strong and rapid sub-flares on timescales as short as $\sim 10$~minutes. The observed luminosity of these sub-flares reached values of $30$~times the spin-down luminosity of the pulsar~\citep{Prsb2017Fermi}. 

In the VHE band, \psrb was detected with the High Energy Stereoscopic System (\hess{}) for the first time in 2004~\citep{psrb_hess_discovery}, after which the array regularly observed the
system at orbital phases close to its periastron passages. VHE observations of \psrb are summarised in~\cite{hess_psrb2020} which reports on five (2004 -- 2017) periastron passages observed with \hess~\citep[see also][for individual analyses of the 2004--2017 periastron passages, respectively]{psrb2004, psrb2007, psrb2011, hess_psrb2020}. See Tab.~\ref{table:obs_details} for specific periastron passage dates and a summary of each passage's VHE observation campaign. 

The VHE light curve obtained from the stacked
analysis of the orbital-period folded data collected during the previous observations of the system indicate the presence of an asymmetric double peak profile~\citep{hess_psrb2020}. Maxima derived from a Bayesian block analysis of stacked data from previous periastron passages were reported between $t_{\mathrm{p}}-32$ and $t_{\mathrm{p}}-26$ days (with a hint of a sub-peak at around $t_{\mathrm{p}} -15$ days) and between $t_{\mathrm{p}}+16$ and $t_{\mathrm{p}}+57$ days, with significances of $12.1 \sigma$ and $39.8 \sigma$, respectively.  

In this work we present the results of the most recent \hess observational campaign on \psrb, performed around the 2021 periastron passage. Extensive coverage of the system during this observation campaign has allowed an unprecedented amount of observational data to be taken post-periastron passage. In particular, observations extended up to the largest post-periastron orbital phase interval in the TeV band to date (29 days more than the previous longest in 2004), see Tab.~\ref{table:obs_details} for details.

Following this introduction, Sec.~\ref{section:Method} outlines the methodology and details of the \hess{} array and its data pipeline. Moreover, this section covers specific details of prior \hess observation campaigns and data analysis of the source during periastron passages up to and including 2021. In Sec.~\ref{Section:Results} the results of the analysis are presented, including studies of the flux behaviour and light curve trends, as well as spectral analysis of the source with a search for spectral variability. In this section we also present our investigation into a correlation between the X-ray / TeV flux and the GeV / TeV flux in the 2021 periastron passage. In Sec.~\ref{Section:Discussion} these results are discussed in the context of previous periastron passages and in the context of the unique findings at other wavelengths in 2021. We also present some theoretical interpretation of the findings of this study. Finally, Sec.~\ref{section:conclusions} contains our concluding remarks.

\section{Method}
\label{section:Method}

\hess{} is an array of five imaging atmospheric Cherenkov telescopes (IACTs), where each telescope is abbreviated and numbered CT1-5, and is located in the Khomas Highlands of Namibia~\citep[see][for detailed descriptions of the \hess{} array]{aharonian_06,hess_psrb2020}.
 
In order to detect Cherenkov light, \hess{} can only be operated under dark conditions. Because of this, \hess{} is not operated during periods of bright moonlight (defined as above $\sim 40\%$ illumination). This results in a cycle-wise data taking period of 28 days. The fundamental data-taking unit of the \hess{} array is an observational run, defined as a period of data acquisition lasting $\sim 28$ mins.

The VHE data presented in this paper are exclusively taken from runs where a minimum of three telescopes from CT1-4 were present (stereo mode). We use  CT1-4 data to allow unbiased direct comparison to the majority of the other periastron passages covered by \hess{}, in which only CT1-4 data was available. For this reason CT5 data were not used in the analysis. The analysis presented in this paper used the reflected regions background method, for light curve production and spectral analysis, as well as the ring background method for the creation of maps \citep[see][for further details of these anaylsis methods]{Berge_07}. Observations were performed using pointing offsets from the source position, all offsets were exclusively performed along right ascension due to the presence of the nearby bright source HESS J1303-631 at an angular separation of $0.75\degree$ North from \psrb. The dataset contained almost exclusively $0.5\degree$ telescope offsets with two runs of 221 (the total run number after data quality selection cuts had been applied) at an offset of 0.7\degree.

Prior to 2021, \hess{} observed \psrb{} covering a number of orbital phases close to previous periastron passages. These include the 2004~\citep{psrb2004}, 2007~\citep{psrb2007}, 2011~\citep{psrb2011}, 2014, and most recently the 2017 periastron passages~\citep[see][for both the 2014 and 2017 periastron passages]{hess_psrb2020}. See Tab.~\ref{table:obs_details} for details of observations in previous periastron passages.   

During the 2021 periastron passage \hess{} attained a total of 100 h of observations in the stereo configuration after data quality selection cuts had been applied. See Tab.~\ref{table:obs_details} for details of the 2021 observations.

\begin{table*}[h]
\centering
\renewcommand{\arraystretch}{1.5}
\begin{tabular}{ c c |c c c c c c} 
 \hline
 & & \textbf{2004} & \textbf{2007} & \textbf{2011} &   \textbf{2014} & \textbf{2017} & \textbf{2021} \\ [0.5ex] 
 \hline
 & Start Date & Feb 27 & Apr 09 & Jan 10 & Mar 07 & Aug 10& Jan 16   \\
 & End Date & Jun 15 & Aug 08 & Jan 16 & Jul 21 & Aug 20 & Jun 16  \\
 & Time $-~t_{\mathrm{p}}$ [days]  & -7 to +98 & -110 to +11 & +26 to +32 & −39 to +78 & −42 to −37 & −23 to +127  \\
\hline
 & $N_{\mathrm{runs}}$ & - & - & - & 141 & 12 & (216) \\
  \textbf{CT5 Mono} & $t_{L}$ $[h]$ & - & - & - & 62.2 & 6 & (99.4) \\
 & $\theta$ $[\degree]$ & - & - & - & 41.8 & 57 & (42.8) \\
 \hline
 & $N_\mathrm{runs}$ & 138 & 213 & 11 & 163 & - & 221 \\ 
  \textbf{CT1-4 Stereo} & $t_{L}$ $[h]$ & 57.1 & 93.9 & 4.8 & 68.1 & - & 100.0 \\
 & $\theta$ $[\degree]$ & 42.5 & 45.1 & 47.6 & 41.9 & - & 42.8 \\

 \hline
\end{tabular}

\caption{Summary of analysed \hess{} observations of \psrb from 2004 to 2021. The number of runs passing spectral quality selection cuts in a given periastron passage is defined as $N_{\mathrm{runs}}$ (see text for further details), $t_{L}$ refers to a periastron passage's total acceptance-corrected observation time, and finally  $\theta$ indicates the mean zenith angle of the periastron passage's observations. Values for years prior to 2021 are taken from \citet{hess_psrb2020} and references therein. CT1-4 stereo data are those runs in which third or more of the CT1-4 telescopes were active. CT5 mono data corresponds to the data obtained from only the central (and largest) telescope CT5. The 2021 CT5 mono observations are displayed in brackets as they are not presented or utilised in this study.}
\label{table:obs_details}
\end{table*}

The results presented in this paper were produced using the HAP (\hess{} Analysis Package)/ImPACT (Image Pixel-wise fit for Atmospheric Cherenkov Telescopes) template-based method chain~\citep{Parsons_2014}. Results have been cross-checked using the Paris Analysis chain~\citep{de_Naurois_2009}.

All light curves and spectra in this work were produced from data that had passed the spectral quality selection cuts, representing the strictest cut criterion for \hess{} data~\citep{aharonian_06}. The data were also subject to a maximum event offset of $2.5^{\circ}$.

To estimate the systematic uncertainties we adopt the values outlined in \citet{aharonian_06} for stereo analyses, as well as compare the reconstructed fluxes and spectral indices between the two major \hess{} analysis chains. This study indicated a systematic uncertainty in the flux at an estimated level of $20\%$ and an uncertainty in spectral indices of $0.1$.
Statistical uncertainties on values/figures in this work (with the exception of spectral parameters that are reported at $95\%$ confidence interval --c.i.--) are given at $68\%$ c.i., unless explicitly stated otherwise.

In calculating the spectra in this study we utilise the forward-folding method~\citep[see][for further information on the forward folding method]{piron_2001}.

\section{Results}
\label{Section:Results}

\begin{figure*}[t]
    \includegraphics[width=0.45\linewidth]{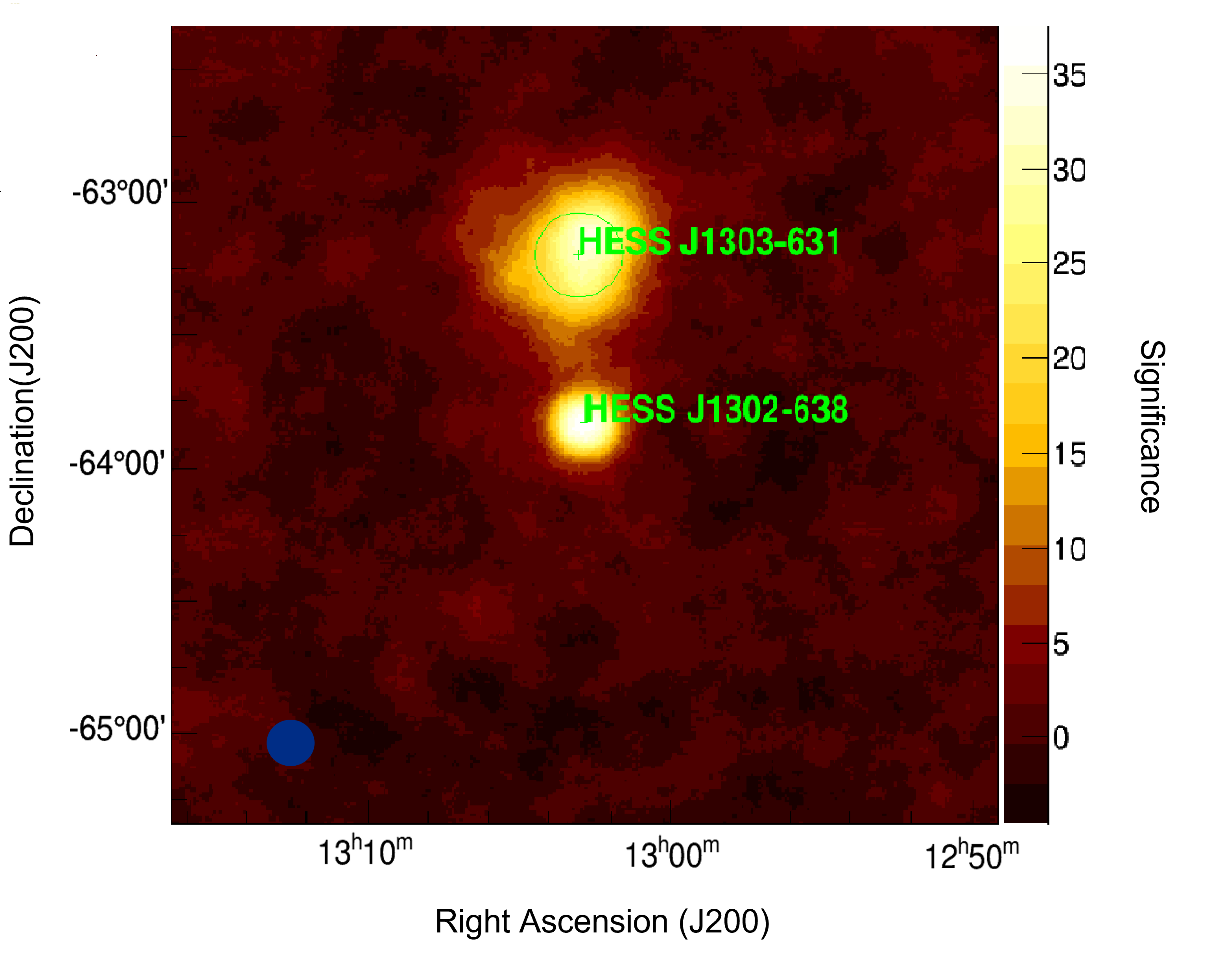}
    \includegraphics[width=0.45\linewidth]{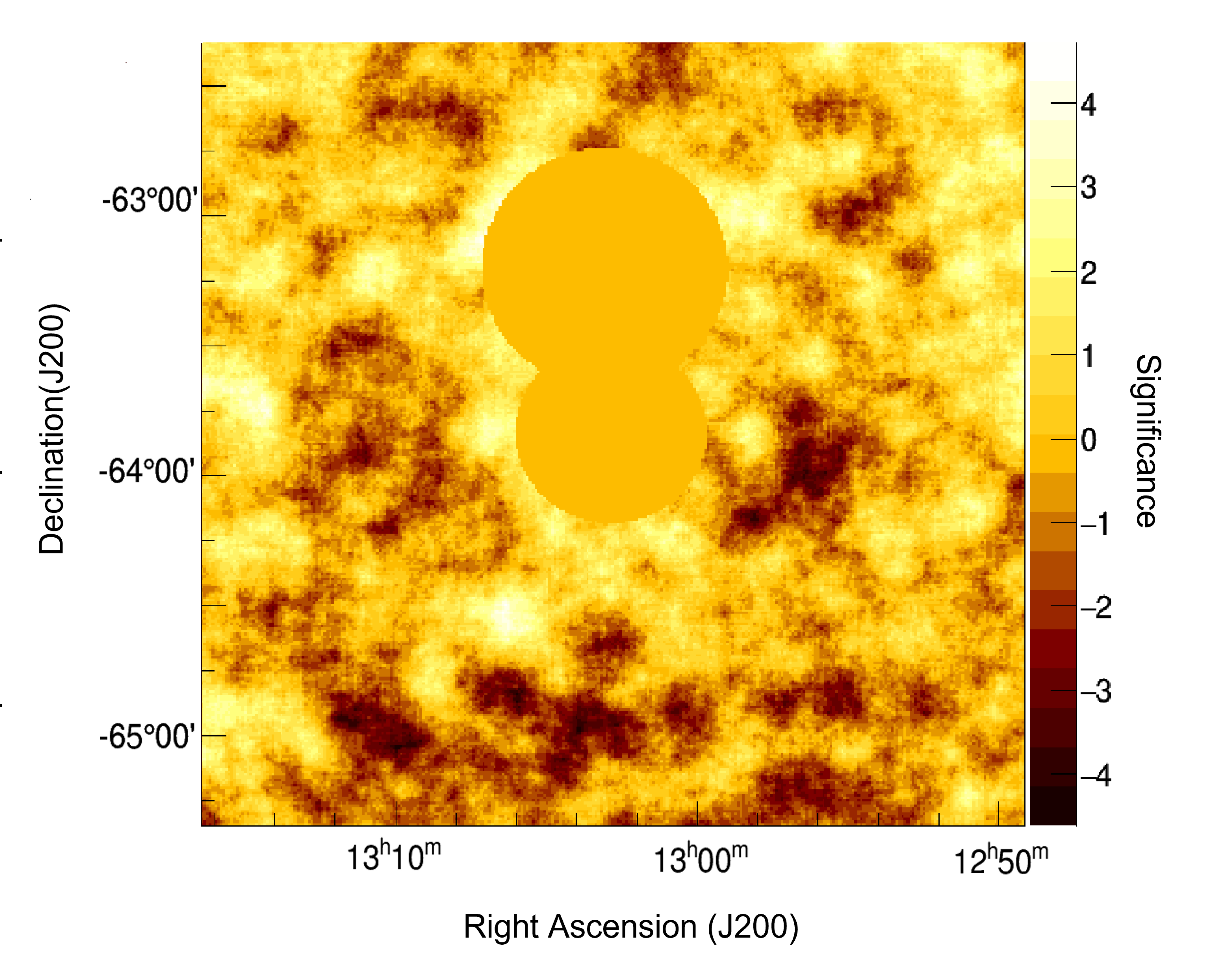}
    \caption{Significance and excluded significance maps. \textbf{Left panel:} VHE significance map displaying PSR~B1259–63/LS~2883 (here labelled as HESS~J1302-638) and the surrounding region. \psrb{} itself is the central object in the image, where also of note is the nearby pulsar wind nebula HESS J1303–631~\citep{psrb2004} directly to the north (see text for further details on this source). Also shown in blue in the lower left of the image, is the 68\% c.i of the point spread function for these observations. \textbf{Right panel:} the significance map of \psrb and its surrounding region after masking the two sources.} The bins in both maps are correlated within a circle of radius $0.1 \degree$. 
    \label{fig:sigmap}
\end{figure*}

The \psrb{} system is located at the J2000 coordinates $RA = 13$h$02$m$47.65$s, 
$Dec = −63 \degree 50' 8.6''$ and is situated in the Galactic plane~\citep{johnston92}. It is near to by the bright source \hessJ{}, a pulsar wind nebula that is spatially coincident with the pulsar PSR~J1301-6305~\citep{hessj2012}. The significance map of the source and its surrounding region are shown in Fig.~\ref{fig:sigmap} using Li and Ma significances \citep{li_ma_1983} and were created by utilising all \hess{} data passing spectral cuts from the 2021 periastron passage ($t_{\mathrm{p}} -23$ days to $t_{\mathrm{p}} +127$ days). \hessJ{} is known to have an energy-dependent morphology with a large spillover at GeV energies \citep{hessj2012,Acero_2013}. This spillover corresponds to an extended and energy dependent emission profile of the source, to a degree that it has the potential to contaminate the emission of nearby sources such as \psrb. This required us to ensure that the effect of spillover was non-existent or negligible at very-high energies by measuring the effect of the spillover in runs far from the periastron passage (using combined data in the period of $\sim t_{\mathrm{p}} +100$ days to $ t_{\mathrm{p}} +500$ days) where VHE emission from \psrb was consistent with zero. No evidence of contaminant emission at VHE energies was found.

During the analysis of \psrb a standard angular distance cut for point sources of $0.005$ deg$^{2}$ was applied (defined as the angular distance between a reconstructed event and the expected source position). 

The background acceptance ratio between the ON and OFF region had a value of $\alpha = 0.07$, resulting in a total excess of $1668.40$ events. In total, for an acceptance corrected live time of $100.02$ hours, we obtain a Li and Ma significance of 36.0 from \psrb. 

\subsection{Spectral analysis}
A full investigation into the VHE spectral properties of the system during the periastron passage was undertaken and several spectra were derived. The total spectrum of the available periastron passage data was calculated, and the spectra of key intervals were created to investigate spectral variability. The first of these time frames included the two \hess{} observational cycles (from $t_{\mathrm{p}} -3.9$ days to $ t_{\mathrm{p}} +15.3$ days) that occurred concurrently with the periastron passage. Secondly, we created a spectrum for the period in which the peak levels of VHE flux were measured (here defined as $t_{\mathrm{p}} +25$ days to $ t_{\mathrm{p}} +36$ days). Additionally, we created a spectrum from the data contemporaneous with the 2021 GeV flare~\citep[here referring to the period $t_{\mathrm{p}} + 55$ days to $ t_{\mathrm{p}} +108$ days as defined in][]{chernyakova_21}. Finally, we created a spectrum of the data from the final two \hess{} observational cycles from $\sim t_{\mathrm{p}} + 81$ days to $\sim t_{\mathrm{p}} +127$ days (from now on referred to as the ``TeV low flux'' period). These datasets will henceforth be referred to as A, B, C, and D, respectively (please refer to Tab.~\ref{table:spectra_details}). Each spectrum of these periods was fit with a power-law model, $\text{d}N/\text{d}E = \phi_{0}(E/E_{0})^{-\Gamma}$, where $\Gamma$ represents the photon index of the power law with a normalisation $\phi_{0}$ and a decorrelation energy of $E_{0}$. The best-fit parameters of these models are presented at $95\%$ c.i. unless otherwise stated. 

To define the energy range for the spectral analysis, two different approaches were used. For the total 2021 spectrum the lower energy bound, at $0.27$ TeV, was defined using Monte Carlo simulations which ensure that the energy reconstruction bias is less than $10 \%$ of the energy~\citep{aharonian_06}. The upper bound of the energy range was defined by the highest energy bin that could be fit with a significance of $2 \sigma$. Henceforth we refer to this energy range as unfixed. 

However, for the total spectrum used for comparison to the sub-periods, and for the sub-periods themselves, a fixed energy range was applied allowing accurate comparison of the different spectra. Thus, we apply an energy fitting range of ($0.4$ - $10.0$) TeV for these sub-periods. The lower bound was chosen such that it supersedes the safe energy threshold for any of the data subsets. The higher energy threshold was chosen to ensure sufficient statistics up to the cut energy for all subsets.

\begin{figure}
    \includegraphics[width=0.99\columnwidth,center]{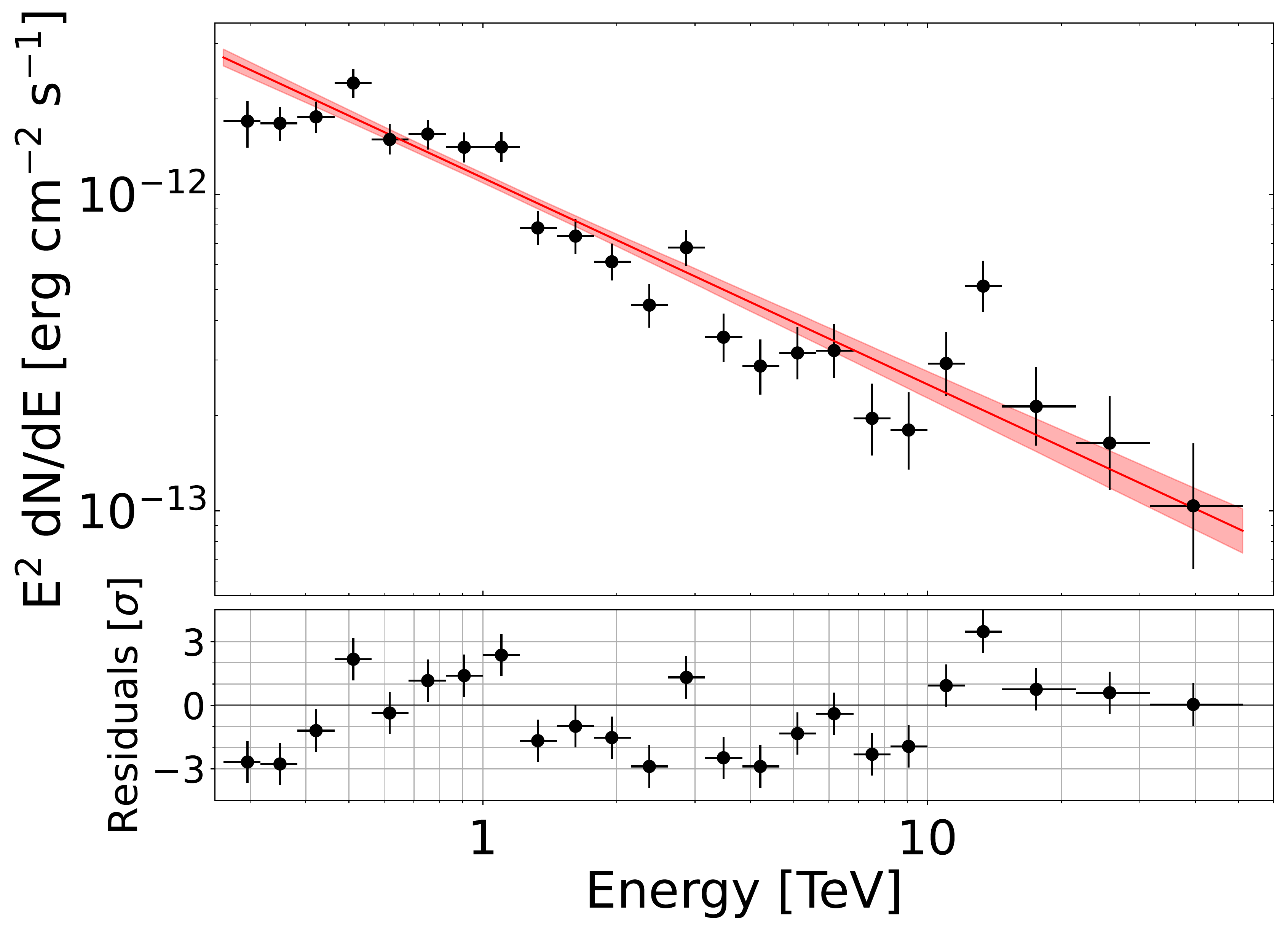}
    \caption{Total spectrum of \psrb's 2021 periastron passage. The spectrum was produced from all data taken with \hess{} during the 2021 periastron passage observations of \psrb, using an unfixed energy range. The spectrum has been fit with a power-law model. For details on the properties of the spectra displayed see Tab.~\ref{table:spectra_details}. The red band indicates the $68 \%$ c.i. of the statistical error for the fitted model. }
    \label{spectra:total}
\end{figure}

Figure~\ref{spectra:total} shows the total 2021 periastron passage spectrum where no pre-fixed energy range for the fit was applied. This spectrum includes all the data taken with \hess{} during the 2021 periastron passage and spans the energy range ($0.30 - 39.6$) TeV (the centres of the lowest and highest energy bins, respectively). Figure~\ref{spectra:total}, shows that the data largely follow a power law, however, there are hints that the spectrum may contain substructure. These substructures could be a result of systematic effects, though an investigation into these effects is beyond the scope of this paper.

\begin{figure}
    \includegraphics[width=0.99\columnwidth,center]{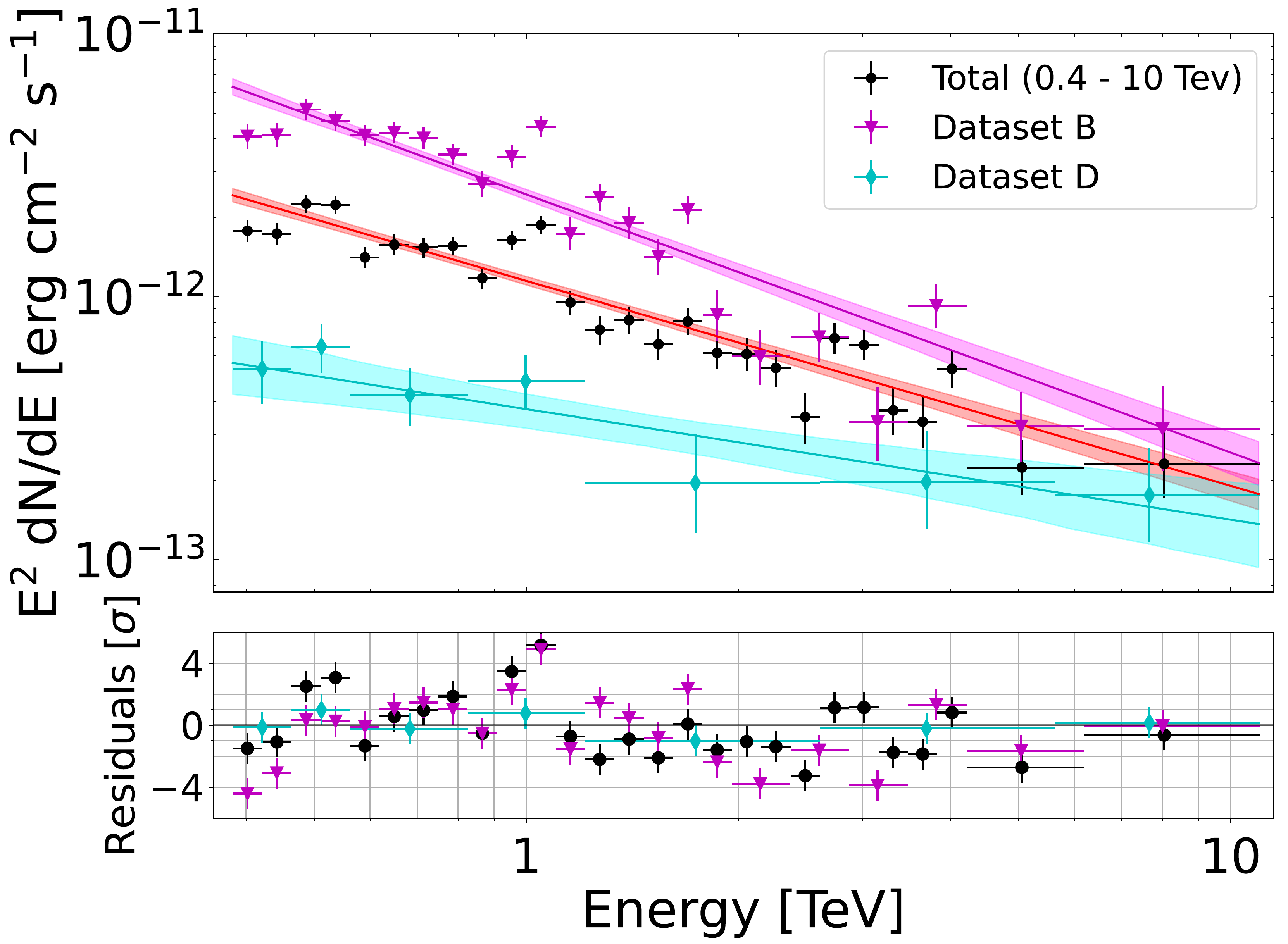}
    \caption{Comparison of spectra from \psrb's 2021 periastron passage. Shown is the total spectrum of the 2021 periastron passage compared to two sub-spectra. Each model was calculated with the forward-folding method, using a power law in an energy range of ($0.4$ - $10.0$) TeV to allow a comparison between them. Comparing spectra allowed us to search for VHE spectral variability on the scale of a single periastron passage. Displayed in red (black circles) is the spectrum of the total 2021 periastron passage, in cyan (cyan diamonds) the spectrum from the dataset D and in magenta (magenta triangles) the spectrum of dataset B. For details on the properties of the spectra displayed see Tab.~\ref{table:spectra_details}. Shaded regions indicate the $68 \%$ c.i. for the fitted model.}
    \label{spectra:comparison}
\end{figure}

Figure~\ref{spectra:comparison} displays two of the three sub-spectra (datasets B and D) created to investigate the spectral behaviour of the system over the course of a single periastron passage. The inclusion of dataset D (the TeV low flux period) allows direct comparison between two unique flux states of the system to search for spectral variability. Additionally the total spectrum of the periastron passage, calculated with a fixed energy range, is shown in this figure. The data points in both Fig.~\ref{spectra:total} and for the sub-spectra in Fig.~\ref{spectra:comparison} were binned to ensure that every flux point has a statistical significance of at least $2 \sigma$.

We find a spectral index for the total periastron passage of $\Gamma = 2.78 ~\pm~ 0.05_{\text{stat}}$ $~\pm~ 0.10_{\text{sys}}$ (for the fixed energy range spectrum) which is consistent with the spectral index of previous years, $\Gamma = 2.76 ~\pm~ 0.03 _{\text{stat}}$ $ ~\pm~ 0.10 _{\text{sys}}$~\citep{hess_psrb2020}. 

We note a statistically significant difference between the spectral indices obtained through the power-law model describing dataset B and D (See Tab.~\ref{table:spectra_details}). Accounting for the uncertainties, the spectral index change between these two datasets is $\Delta \Gamma = 0.56 ~\pm~ 0.18_{\text{stat}}$ $~\pm~0.10_{\text{sys}}$, implying a sub-orbital spectral variation at a c.i. of greater than 95\%. 

For the total unfixed spectrum, we attempted to fit an exponentially cut-off power-law model in order to determine if the data shows a preference for a high-energy cut-off. This revealed that a model with a cutoff is not preferred, with a lower limit on the cut-off energy of $E_{C}^{95\%} = 27.1$ TeV.

\begin{table*}
\centering
\renewcommand{\arraystretch}{1.6}
\begin{tabular}{ p{3.5cm} | p{2.8cm} | p{3.8cm}| p{3.8cm} | p{2.2cm} } 
 \hline
 \textbf{Dataset} & \textbf{Time $-~t_{\mathrm{p}}$} &\textbf{Spectral Index ($\Gamma$)} & \textbf{Normalisation ($\phi_{0}$)} \newline Decorrelation \newline Energy ($E_{0}$)  &  \textbf{Normalisation ($\phi_{0}$)} \newline (1 TeV) \\
  & days & & $10^{-12}$ TeV$^{-1}$  cm$^{-2}$     s$^{-1}$ & $10^{-12}$ TeV$^{-1}$  cm$^{-2}$     s$^{-1}$ \\
  \hline
  Total 2021 & $-23.58$ to $+127.26$ & $2.65~ \pm~ 0.04 _{\text{stat}}$ $ ~\pm~ 0.10 _{\text{sys}}$ & $1.28 ~\pm~ ´0.05_{\text{stat}}  ~\pm~  0.26 _{\text{sys}}$  & $1.13 ~\pm~ ´0.04_{\text{stat}} \newline  ~\pm~  0.23 _{\text{sys}}$  \\
  ($0.27$ - $33.6$ TeV) &&& (0.95 TeV) \\
  \hline
  Total 2021 & $-23.58$ to $+127.26$ & $2.78 ~\pm~ 0.05 _{\text{stat}}$ $ ~\pm~ 0.10 _{\text{sys}}$ & $1.31 ~\pm~ ´0.05_{\text{stat}}  ~\pm~  0.26 _{\text{sys}}$ &  $1.15 ~\pm~ ´0.04_{\text{stat}} \newline ~\pm~  0.23 _{\text{sys}}$ \\
  ($0.4$ - $10$ TeV) &&& (0.95 TeV)\\
 \hline
 2021 Periastron Period \newline (\textit{Dataset A}) & $-3.9$ to $+15.3$ & $2.75 ~\pm~ 0.11 _{\text{stat}}$ $ ~\pm~ 0.10 _{\text{sys}}$ & $1.52 ~\pm~ ´0.12_{\text{stat}}  ~\pm~  0.30 _{\text{sys}}$ &  $1.34 ~\pm~ ´0.10_{\text{stat}} \newline ~\pm~  0.27 _{\text{sys}}$ \\
  ($0.4$ - $10$ TeV) &&& (0.95 TeV)\\
 \hline
 2021 Peak TeV Flux \newline (\textit{Dataset B}) & $+25$ to $+36$ & $2.98 ~\pm~ 0.07 _{\text{stat}}$ $ ~\pm~ 0.10_{\text{sys}}$  & $5.00 ~\pm~ 0.22_{\text{stat}} ~\pm~ 1.00 _{\text{sys}}$ &  $2.45 ~\pm~ 0.12_{\text{stat}} \newline ~\pm~ 0.49 _{\text{sys}}$ \\ 
 ($0.4$ - $10$ TeV) &&& (0.79 TeV) \\
 \hline
  2021 GeV Flare Period \newline (\textit{Dataset C})& $+55$ to $+108$ & $2.42 ~\pm~ 0.10 _{\text{stat}}$ $ ~\pm~ 0.10 _{\text{sys}}$  & $0.38 ~\pm~ 0.03_{\text{stat}} ~\pm~ 0.08 _{\text{sys}}$ & $0.68 ~\pm~ 0.06_{\text{stat}} \newline ~\pm~ 0.14 _{\text{sys}}$ \\ 
  ($0.4$ - $10$ TeV) &&& (1.27 TeV) \\
 \hline
   TeV Low Flux Period \newline (\textit{Dataset D})& $+81$ to $+127$& $2.42 ~\pm~ 0.17 _{\text{stat}}$ $ ~\pm~ 0.10 _{\text{sys}}$ & $0.17 ~\pm~ 0.02 _{\text{stat}}$ $ ~\pm~ 0.03 _{\text{sys}}$ & $0.37 ~\pm~ 0.05 _{\text{stat}}$ \newline $ ~\pm~ 0.07 _{\text{sys}}$ \\
   ($0.4$ - $10$ TeV) &&& (1.40 TeV)\\
 \hline
  Average of Previous Periastron Passages & $-106$ to $+98$& $2.76 ~\pm~ 0.09 _{\text{stat}}$ $ ~\pm~ 0.10 _{\text{sys}}$ & - & $1.55 ~\pm~ 0.12 _{\text{stat}}$ $ ~\pm~ 0.31 _{\text{sys}}$ \\
 \hline
\end{tabular}
\caption{Comparison of the best-fitting spectral parameters when a power-law model is fit to different datasets within the 2021 periastron passage of \psrb. $\Gamma$ represents the best-fit photon index of the power law $\text{d}N/\text{d}E = \phi_{0} (E/E_{0})^{-\Gamma}$, with $\phi_{0}$ denoting the best-fit normalisation level, and $E_{0}$ being the decorrelation energy~\citep{Abdo_2009}. Presented here are the spectral properties of the total dataset of the periastron passage (for both fixed and unfixed energy ranges), as well as four additional sub-orbital periods from the 2021 dataset. All 2021 spectra in this table (with the exception of the total periastron passage spectrum marked ``unfixed energy range'') are calculated using an energy range of ($0.4$ - $10.0$) TeV to enable comparison. The average spectral properties of \hess{} stereo analysis from previous periastron passages~\citep[2004, 2007, 2011 and 2014, taken from][]{hess_psrb2020} are also included for comparison, from an averaging of the values reported in these papers. The spectra of previous periastron passages, however, do not use the same fixed energy values as the 2021 data. The two errors associated with each value are representative of, first, the magnitude of the statistical uncertainty in the value at a $95 \%$ confidence level, and secondly the systematic error in the measurements as adopted from \cite{aharonian_06}.}
\label{table:spectra_details}
\end{table*}

\subsection{Flux analysis}

\begin{figure*}[t]
    \centering
    \includegraphics[width=\textwidth, height=\textheight, keepaspectratio]{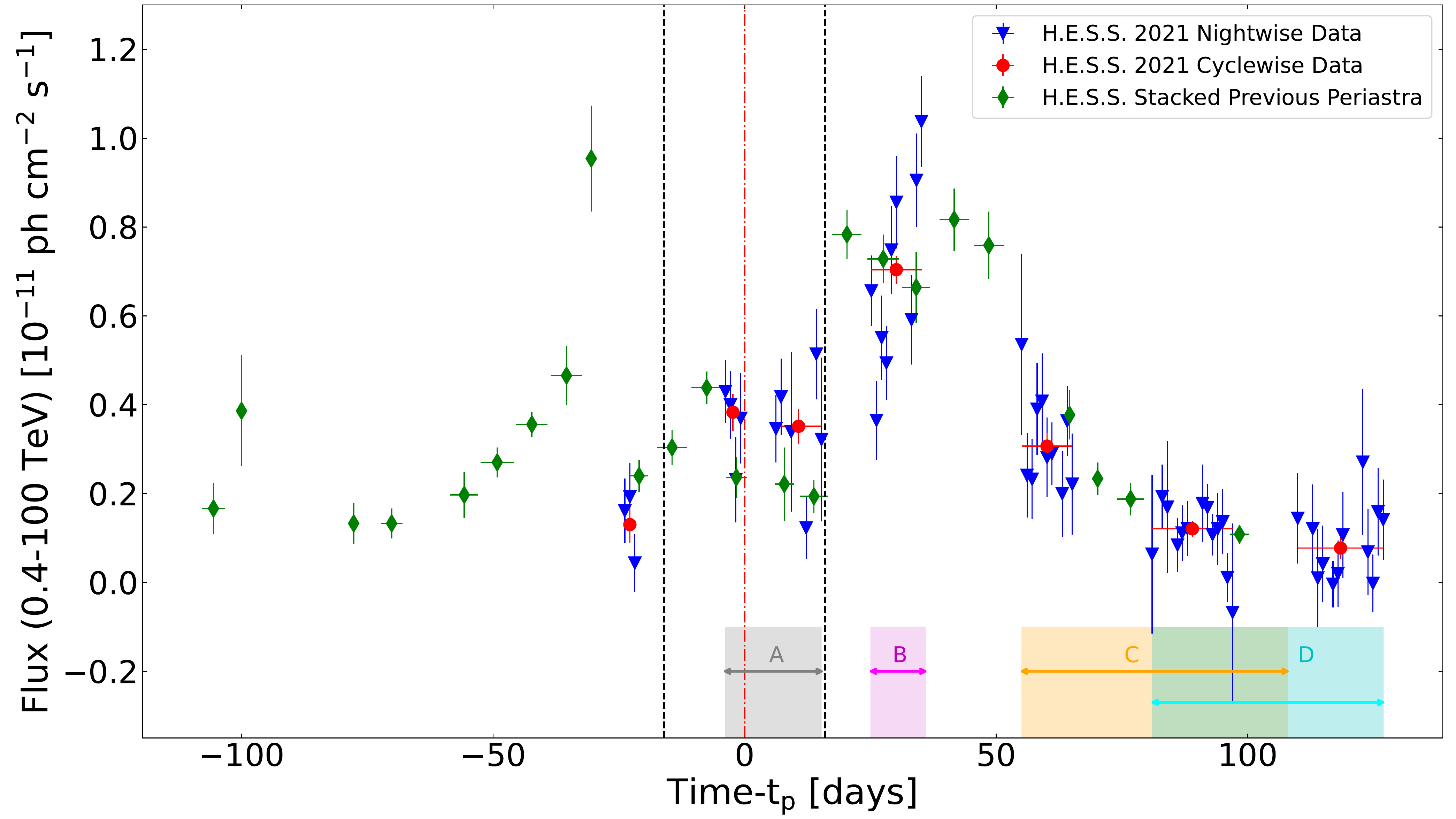}
    \caption{VHE light curve of \psrb{}'s 2021 periastron passage. VHE Data is shown in different binnings:  night-wise binned fluxes (blue triangles), cycle-wise binned fluxes (red dots), and stacked data from \hess{} observations of previous periastron passages (green diamonds). Horizontal error bars correspond to times of the earliest and latest runs that were merged to make the data point. Cycle-wise data contain the merged runs within one observational cycle ($\sim 28$ days). Stacked data points have a weekly binning and correspond to data from the 2004, 2007, 2011, 2014 and 2017 periastron passages; these points were taken from \cite{hess_psrb2020}.  
    The two dashed black lines at $t_{\mathrm{p}} -16$ days and $t_{\mathrm{p}} +16$ days correspond to the time at which the disc crossing is thought to occur, the dot-dashed red line at $t_{\mathrm{p}} = 0$ days represents the point of periastron in the system. Finally, the shaded areas with arrows are displayed as a visual representation of the time periods from which sub datasets were taken. These regions are also marked with the letters of the dataset they represent. The grey region represents the time period encompassed by dataset A, the magenta region corresponds to the time period of dataset B, and the yellow region and the cyan regions indicate the time periods of datasets C and D, respectively; see Tab.~\ref{table:spectra_details} for the full details of these sub-periods.}
    \label{fig:lc}
\end{figure*}

For the 2021 periastron passage dataset, light curves were produced for the \hess{} data in two different integration timescales, see Fig.~\ref{fig:lc}. These were: night-wise binning and cycle-wise binning -- grouping runs to individual observational cycles of $\sim 28$ days.

Individual flux points and their uncertainties were calculated using a reference spectrum, in this work this was a a power-law model in the energy range ($0.4$ - $100.0$) TeV. An index of $\Gamma = 2.65$ was used, corresponding to the total 2021 dataset spectral index value.

We performed a search for variability in \psrb over a number of different timescales, however, statistical uncertainties prevented us from establishing the presence of variability on run-to-run timescales\footnote{Most commonly, subsequent runs were taken during the same or the following
night, with the exception of several breaks due to moonlight or bad
weather. Thus, run-to-run timescales range from a few hours to a few
days.}. Our analysis of the VHE data at $25-35$~days after periastron (the time period of fastest VHE flux increase) indicates that a model with a linearly increasing flux is a better fit to the data in this period than a constant flux model. This was determined by comparing the chi-squared values of a linear increase model, and a constant flux model, in this time period. The comparison of the fits of these methods showed that a linear flux increase is preferred at a $\sim 4\sigma$ level. During this time period the flux increased by a factor of two. Other than this increase in a period of $\sim 10$~days, we did not find significant evidence for a linear flux increase at shorter timescales. Thus, we see variability on timescales of  down to $\sim 10$ days. It is possible that there exists variability on shorter timescales, however, we are unable to probe this due to statistics.

We investigated the impact of using an assumed spectral index of $\Gamma = 2.65$ to calculate the night-wise fluxes, given the discovery of sub-orbital spectral index variation. We investigated this by calculating binned fluxes using the two extreme values of the spectral index $\Gamma = 2.42$ (from the spectrum of the emission from dataset D) and $\Gamma = 2.98$ (dataset B). We then evaluated the difference between the nightwise fluxes of the two light curves that these indices produced. The percentage difference between the flux of the two new light curves yielded a maximum systematic error in the flux of $\pm ~10\%$, a comparable value to that of \citet{aharonian_06} from which the systematic error values of this study were taken (see Sec.~\ref{section:Method} for details). This value represents an additional systematic flux error in the light curves exclusively, and does not have an impact on any scientific conclusions in the paper.

Although the 2021 VHE light curve presented in Fig.~\ref{fig:lc} shows an overall trend similar to the light curve obtained from the stacked analysis of the orbital-period-folded data collected during previous observations~\citep{hess_psrb2020}, we argue that a detailed comparison of the system's flux behaviour is complicated by the different coverage of the \hess datasets. Despite observing at orbital phases close to the second disc crossing in the 2021 dataset, we do not see a VHE flux enhancement around this time. However, we report a VHE maximum occurring between $t_{\mathrm{p}}+20$ and $t_{\mathrm{p}}+50$ days~\citep[seen during the period of the maximum reported in previous years,] []{hess_psrb2020}.

\subsection{2021 GeV flare}

The 2021 GeV flare (shown in Fig.~\ref{fig:lightcurve _correlated}) differed in considerable ways from those of previous periastron passages (although the GeV behaviour appears inherently variable between periastron passages). As in 2017, the 2021 GeV flare started at $\sim t_{\mathrm{p}} +55$ days with GeV activity extending to $\sim t_{\mathrm{p}} +108$ days, see \citet{chernyakova_21}.  The system underwent numerous rapid and energetic sub flares on very short timescales (in some cases as short as $\sim 10$ minutes) reaching up to 30 times the spin-down luminosity of the pulsar.

We see no correspondence to the 2021 GeV flare, from $\sim t_{\mathrm{p}} + 55$ days to $\sim t_{\mathrm{p}} +108$ days, in the VHE light curve (see Sec.~\ref{sec:results:gev_tev_correlation} for further investigation into this). We do, however, note that our ability to monitor this is somewhat complicated by a large gap in our observations during the 2021 GeV flare period, as no observations were performed in the time frame of $\sim t_{\mathrm{p}} + 65$ days to $\sim t_{\mathrm{p}} +81$ days.

The spectrum of dataset C (derived during the 2021 GeV flare period) has a spectral index notably similar to that of dataset D. There is, therefore, also a discrepancy in the spectral index between datasets B and C at $\Delta \Gamma = 0.56 ~\pm~ 0.18$, the same level as the previously discussed discrepancy between dataset B and D.

\subsection{X-ray-TeV correlation}

\begin{figure*}[]
    \includegraphics[width=1.02\columnwidth]{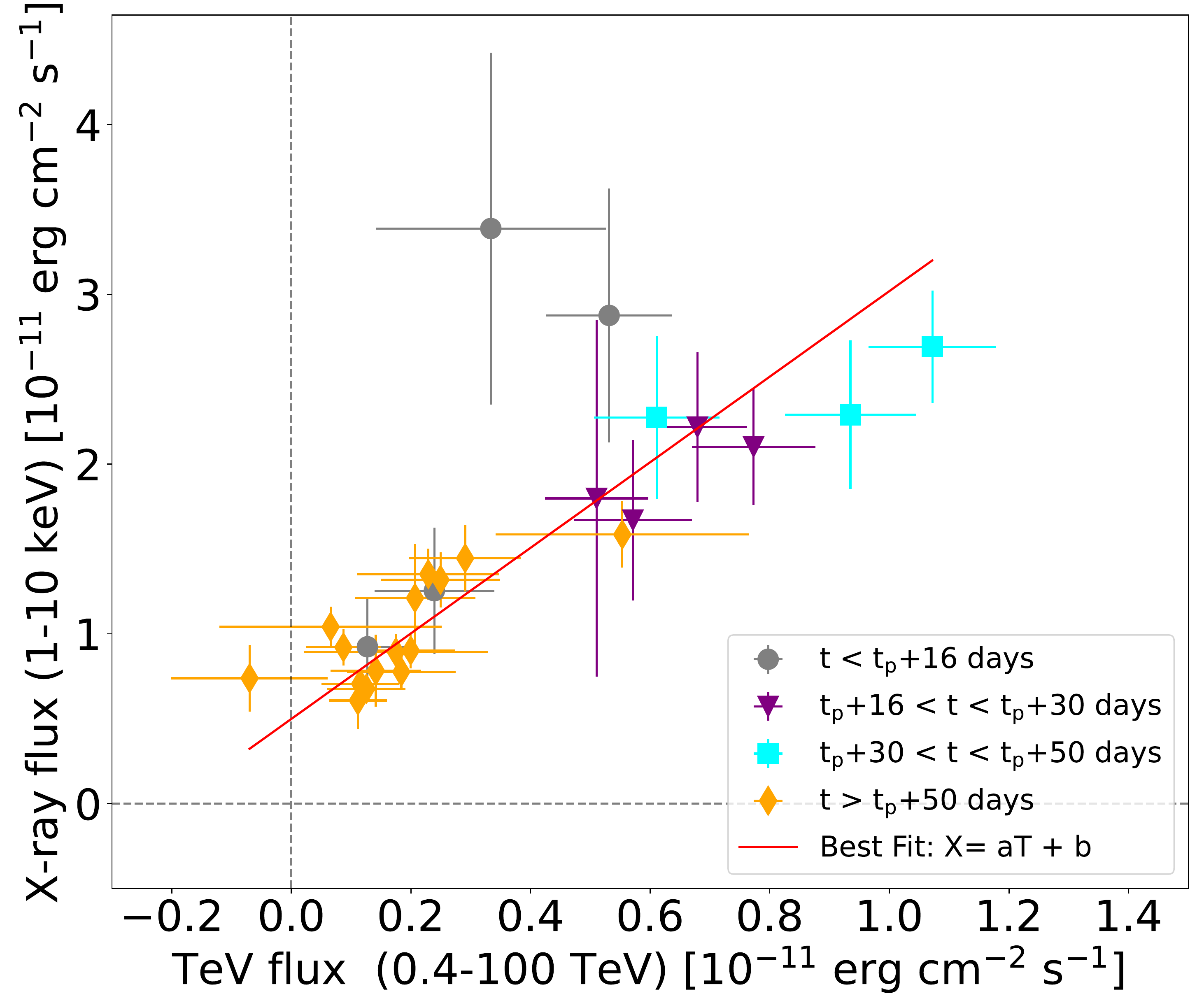}
    \includegraphics[width=1.035\columnwidth]{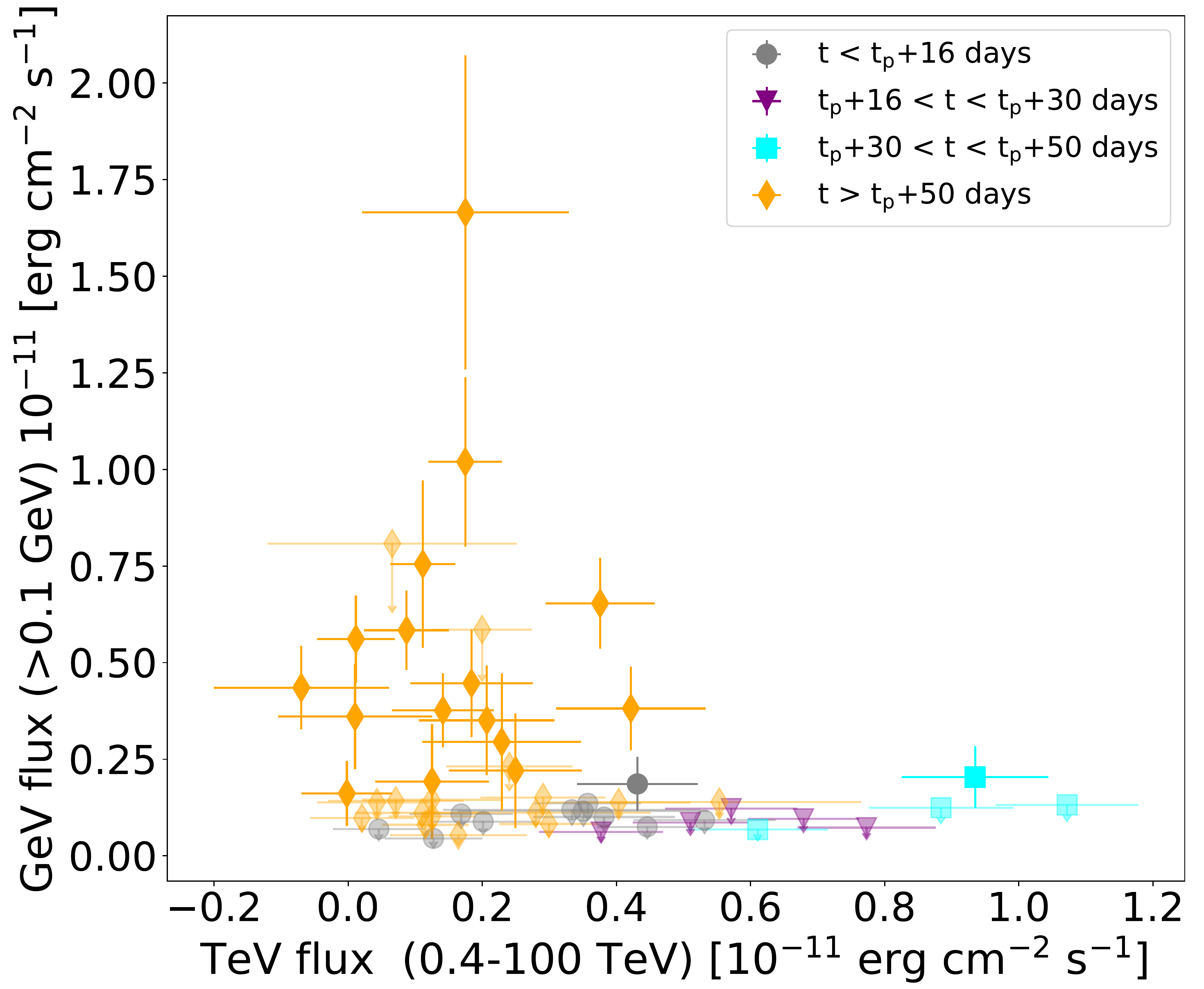}
    \caption{
     Linear correlations between X-ray-TeV and GeV-TeV data sets. \textbf{Left panel:} Linear correlation ($F_{X} = aF_{TeV} + b$) between the flux of the TeV and X-ray datasets, where exclusively integrated energy flux points measured within a day of one another were utilised. Shown in red is the line of best-fit resulting from a linear fit with an offset. A large outlier can be seen at TeV flux $\sim 0.3 \times 10^{-11}$ erg cm$^{-2}$ s$^{-1}$; this outlier originates from rapid X-ray variability in the period from periastron up to $\sim t_{\mathrm{p}} +16$ days. The dashed lines mark the zero points of each axis. As can be seen, there is a non-zero X-ray flux value at zero TeV flux in the fit.\newline
    \textbf{Right panel:} Correlation plot of the flux of the TeV and GeV datasets, where exclusively flux points measured within a day of one another were used. Note that transparent points correspond to upper limits taken from the \flat daily light curve, where each upper limit has been assumed to have a flux of half the upper limit value and a flux error corresponding to a $95\%$ c.i. on the \textit{Fermi} upper limit. GeV flux points have been adapted from \cite{chernyakova_21}. In both figures the orbital phases from which correlation points are taken are denoted by their colour (see Fig.~\ref{fig:lightcurve _correlated}).}
    \label{fig:Xray_TeV_linear}
\end{figure*}

We investigated a potential correlation between VHE and X-ray flux in the 2021 periastron passage data. In this study we utilise the results reported in~\citet{chernyakova_21}, from the Neil Gehrels \textit{Swift} (\textit{Swift}) X-ray telescope (XRT) and the Neutron Star Interior Composition Explorer \textit{NICER} (both in the $0.3 - 10$~keV range), covering the time period January, 19th, 2021 to May, 24th, 2021 ($t_{\mathrm{p}}-21$ to $t_{\mathrm{p}}+103$~days).

In order to perform the correlation study on timescales relevant to the system's behaviour, we binned the data from \hess{} on a nightly basis, resulting in a total of 57 TeV points, compared to a total of 96 X-ray points binned by observations (32 of which from \textit{NICER}, 64 from \textit{Swift}). X-ray points had an average separation in time of $1.24$ days (excluding time gaps of greater than a week) over the whole dataset.

A sub-selection of observations from both the TeV and X-ray datasets was made, ensuring the selection of only X-ray points occurring within one day of a TeV point. This correlation timescale of a day was selected because this is the shortest timescale in which the available statistics could confirm a lack of variability of \psrb in X-rays. To make this sub-selection, we iterated over the data in steps of correlation timescale, where any TeV and X-ray points within this time became a correlated pair. Instances where two or more points of the same data type (X-ray or TeV\footnote{Formally, due to the TeV data being binned on a nightly basis it was not possible for two runs  to occur within the same correlation timescale of a day. Thus this step only affected X-ray data points, which frequently occurred within a day of other X-ray points}) were found to be within a single correlation timescale were handled by averaging the flux, time and uncertainties of the respective points. Any data points that did not contain a counterpart within one day of the point were not considered in the correlation study.

After this selection a total of 26 correlated pairs were found. This selection of correlated pairs and their distribution across the periastron passage can be seen in Fig.~\ref{fig:lightcurve _correlated}. The first correlated pair is at a time of $t_{\mathrm{p}} -1.53$ days, extending up to the time of the final pair at $t_{\mathrm{p}} +97.47$ days. The majority of pairs occurred at times later than $t_{\mathrm{p}} +50$ days.   

Figure~\ref{fig:Xray_TeV_linear} shows the results of the correlation investigation between the two datasets. By minimising $\eta^2$ (where $\eta^{2}$ is a linear combination of $\chi^{2}$ tests, see appendix~\ref{appendix:eta} for details of the $\eta^{2}$ test) we obtained the following best-fitting values of the linear fit parameters for the model $F_{X} = aF_{TeV} + b$: $a = 2.62 ^{+0.41}_{-0.38}$ and $b = 0.50^{+0.12}_{-0.13} \times 10^{-11}$~erg cm$^{-2}$ s$^{-1}$ (where $F_{X}$ and $F_{TeV}$ are the X-ray and TeV fluxes, respectively). These values of the fit gave a total $\eta^{2} = 105.08$ for 26 correlated pairs, resulting in $\Bar{\eta}^2 = 2.10$.

To estimate the statistical uncertainties we performed numerical simulations, considering $N = 10^{6}$ random trial datasets. The integrated X-ray and TeV photon flux for each trial dataset were simulated from the original data, assuming a Gaussian distribution of uncertainties. The quoted errors for each parameter correspond to a 68\% c.i. of all best-fit values obtained during random trial datasets when fit with the same model. We estimated the chance probability of finding a correlation at $\Bar{\eta}^2 = 2.10$ by comparing the number of trials that provided better $\eta^2$ values than the original data, to the number of trials with a worse $\eta^2$. Making this comparison we found a chance probability of $2.27\times 10^{-3}$.

We therefore conclude that there is a positive correlation between the X-ray and TeV flux during the time periods of the 2021 perisatron that were probed by the study. While all points follow this linear trend, there are two notable outliers from the correlation. These pairs represent X-ray and TeV flux points that were measured shortly before $t_{\mathrm{p}}+16$~days (see Figs.~\ref{fig:lightcurve _correlated} and \ref{fig:Xray_TeV_linear}). With regards to the conclusion of a linear correlation, it is important to note that this initial study, which placed no restrictions on the time of the flux points used in the correlation, consisted mostly of flux points that were measured at times greater than $t_{\mathrm{p}}+25$~days (see Fig.~\ref{fig:lightcurve _correlated}). This uneven sampling across the periastron passage prevents us from establishing the presence of such a correlation before this time period.

We separately assessed the correspondence of the VHE flux level to either the third X-ray peak or to the gradual decay seen in the X-ray flux profiles of previous years. To achieve this, we performed model fitting on the full dataset of 2021 VHE flux data, separately fitting both a negative exponential function and a negative exponential function combined with a Gaussian. Here, the negative exponential model is representative of the X-ray flux behaviour  (corresponding well to the behaviour seen in previous years) and the Gaussian represents the flux profile of the third X-ray peak in 2021. In this comparison, the VHE data was better fit by the negative exponential function summed with a Gaussian, at a $5.5 \sigma$ level. However, this fit is based on the available VHE data points that occur only immediately before the peak (see Fig.~\ref{fig:lightcurve _correlated}). Therefore, the limited number of points immediately after the peak makes it difficult to draw more robust claims.

In addition, we undertook a separate correlation study utilising only data occurring after the time of the second X-ray peak in 2021 ($t_{\mathrm{p}} +16$ days). The data from before this point, notably, were the cause of the blue coloured outliers at greater than $68\%$ c.i. seen in the linear correlation in Fig.~\ref{fig:Xray_TeV_linear}. We apply exactly the same method as described previously. This results in a goodness-of-fit value of $\eta^{2} = 57.77$ that, for 22 correlated pairs, results in a $\Bar{\eta}^2 = 1.38$ (compared to a value of $\Bar{\eta}^2 = 2.10$ for the unrestricted dataset) and gave a  chance probability of $8.3 \times 10^{-4}$.

\subsection{GeV -- TeV correlation}
\label{sec:results:gev_tev_correlation}

We also conducted a study into the correlation between the 2021 TeV and GeV datasets to further quantify the apparent absence of a TeV counterpart to the 2021 GeV flare. We utilise daily-binned GeV flux data from \cite{chernyakova_21} for comparison with the nightly-binned TeV data. The method of correlation used was identical to that of the previous study between X-ray and TeV data. Despite known sub-flares at GeV energies, sometimes on timescales of ten minutes, we opted to utilise a timescale of one day. This correlation length matched the binning of the two light curves, and is also consistent with the X-ray -- TeV correlation results.

Following the correlated pair selection, a total of 56 correlated pairs (from 57 TeV points and 187 GeV points) were selected. The first of these is at a time of $t_{\mathrm{p}} -23.5$ days, extending up to $t_{\mathrm{p}} +126.5$ days. Because all but one TeV points are time-correlated with a GeV point, in the GeV/TeV correlation these pairs are relatively evenly sampled across the time range of the periastron passage, reflecting the distribution of the TeV points.
The majority of the correlated GeV points, however, are upper limits from the \textit{Fermi} analysis (144 of 187 points). We therefore tested for a correlation using several approaches. Firstly, we simply omitted the upper limits from the correlation study. This resulted in an $\Bar{\eta}^2 = 11.56$. However such a selection introduces a bias towards high GeV fluxes and could mask an existing correlation. We therefore examined two approaches including upper limits, these methods corresponded to adopting a Gaussian distribution as the probability density function (PDF) for the upper limits \citep[see][for further details and alternative treatments of the PDF]{kelly_2007}. The first of these approaches was to utilise half the upper limit value as the flux value and a dispersion corresponding to $95\%$ c.i. of the \textit{Fermi} upper limits. The second method utilised a zero value as the flux estimate (and once again a dispersion corresponding to $95\%$ c.i. of the \textit{Fermi} upper limits). The resulting $\Bar{\eta}^2$ values were $15.02$ and $10.67$ for the Gaussian centered on half the upper limit value and the Gaussian centered at zero, respectively. All of these tests excluded a correlation at levels greater than $5 \sigma$. We conclude that (within the uncertainties of the measurements) we detect no significant GeV -- TeV correlation throughout the entire probed time period.

\section{Discussion}
\label{Section:Discussion}
Understanding the broadband emission from GRLBs is a complex problem, which still awaits a definite solution. Despite the difficulties, some progress, however, has been made in the modeling of emission from these systems.  
\psrb contains a non-accreting pulsar, thus, in what follows we discuss the properties of the emission in the framework of a binary pulsar model.
This scenario implies that the relativistic outflow from a rotation powered pulsar interacts with the stellar wind which, in the case of \psrb, consists of a radiation-driven polar wind, and a significantly more dense Keplerian-like decretion disc.

\afterpage{
\begin{landscape}

\begin{figure}[]
\centering
\includegraphics[width=0.96\linewidth, height=0.8\textheight, keepaspectratio]{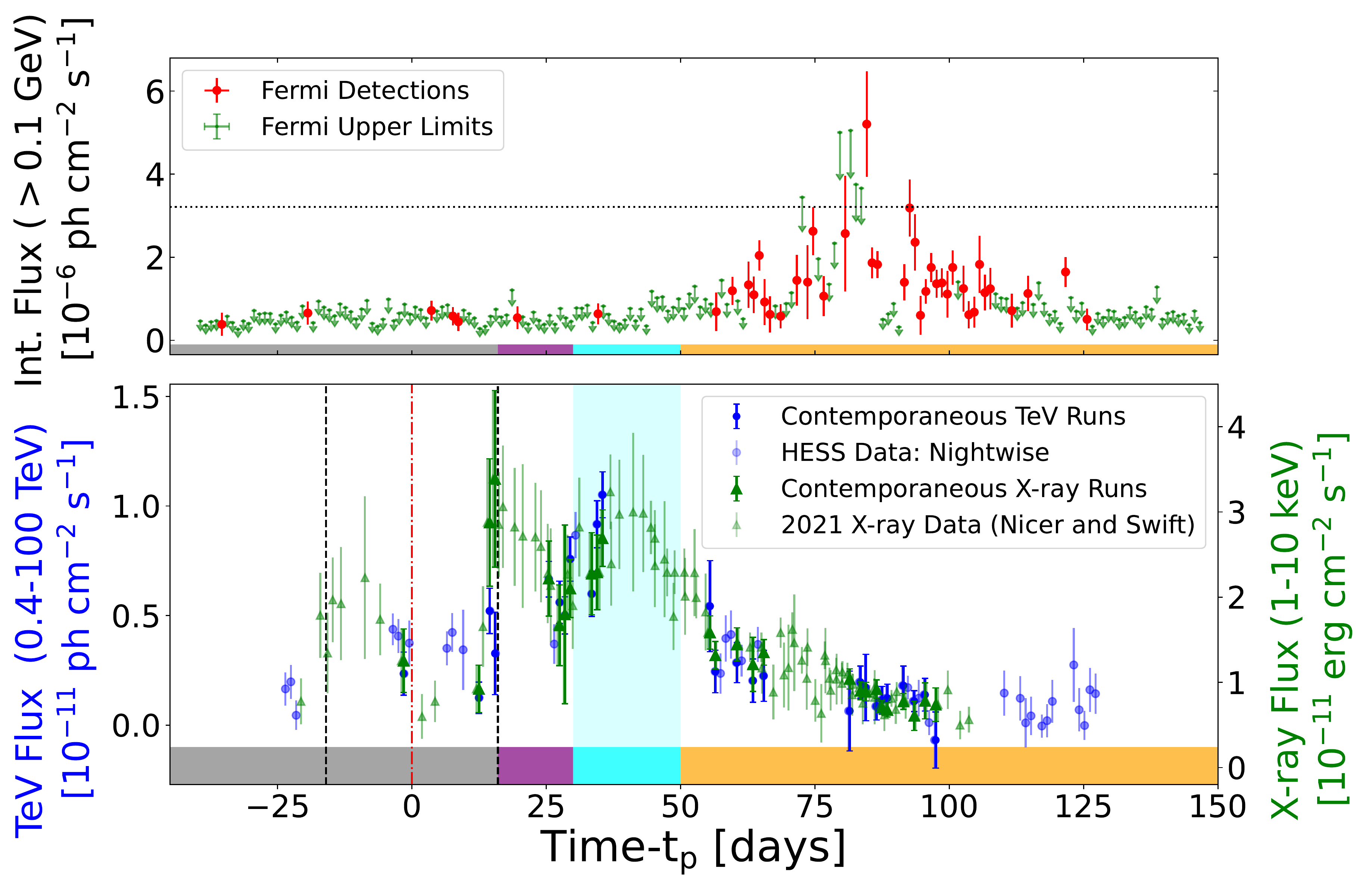}

    \caption{Comparison of the 2021 GeV light curve to the correlated X-ray and VHE light curves. \textbf{Upper panel:} The GeV light curve of \psrb{}'s 2021 periastron passage observed with \flat ($0.1 - 10$ GeV). Figure adapted from \citet{chernyakova_21}. Data points represent daily binnings, with green points indicating $95\%$ upper limits and red points denoting detections ($TS > 4$). The horizontal black dashed line marks the flux value corresponding to the spin-down luminosity of \psrb's pulsar $L_{\mathrm{sd}} = 8.2 \times 10^{35}$ erg s$^{-1}$ \citep[see][]{chernyakova_21}. \textbf{Lower panel:} Light curve of \psrb{}'s 2021 periastron passage displaying both VHE flux points from \hess{} and X-ray data from the \textit{Swift} and \textit{NICER} observatories. Points with no transparency are points that were selected by the time correlation step and fell within a day of an alternate type of observation (see text for further details). Translucent points are data from each respective dataset that did not pass the time correlation step. The lines at $\sim t_{\mathrm{p}} ~\pm~ 16$ days represent the assumed disc crossing times and the line at $t_{\mathrm{p}} = 0$ days marks the time of periastron. 
    In both panels, coloured regions correspond to the orbital phases of correlation points in Fig.~\ref{fig:Xray_TeV_linear} that share the same colour. Here the grey region represents orbital phases before $t_{\mathrm{p}} +16$ days, the purple region represents the time frame $t_{\mathrm{p}} +16$ to $t_{\mathrm{p}} +30$ days, the cyan region is the time frame $t_{\mathrm{p}} +30$ to $t_{\mathrm{p}} +50$ days (coincident with the third X-ray peak reported in 2021 X-ray data), and the orange region corresponds to any time after $t_{\mathrm{p}} +50$ days. }
    \label{fig:lightcurve _correlated}
\end{figure}

\end{landscape}
}

\subsection{Orbital dependence of the X-ray and TeV emission}
The termination of the pulsar wind occurs at a distance of $R_{\mathrm{ts}}$ (from the pulsar) where the ram pressure of the stellar and pulsar winds are equal.
Given the much higher anticipated speed of the pulsar wind, the energy injection of non-thermal particles into the interaction region is dominated by contributions from the pulsar.
Therefore, one expects that the radiation processes in binary pulsar systems are similar to those taking place in pulsar wind nebulae \citep{1994ApJ...433L..37T,tavani_1997}, however, with the caveat of some important modifications.

Firstly, as the magnetic field is provided by the pulsar wind, a smaller termination distance necessarily implies a stronger magnetic field:
\begin{equation}
  \begin{split}
    B \lesssim B_{\textrm{max}}&= \sqrt{\frac{L_{\mathrm{sd}}}{c R_{\mathrm{ts}}^2}}\\
                        &\approx 3 \left(\frac{L_{\mathrm{sd}}}{8.2\times10^{35}\ergs}\right)^{1/2}\left(\frac{R_\mathrm{ts}}{0.1\mathrm{AU}}\right)^{-1}\,\mathrm{G}\,,
  \end{split}
\end{equation}
where $L_{\mathrm{sd}}$ is the pulsar's spin down luminosity. The second important difference is that the photon field is dominated by contributions from the optical companion. This provides an intense photon field with an energy density of
\begin{equation}
  \begin{split}
    w_{\mathrm{ph}}&= \frac{L_*}{4\pi c  R^2}\\
                   &\approx 3\left(\frac{L_*}{2.3\times10^{38}\ergs}\right)\left(\frac{R}{1\,\mathrm{AU}}\right)^{-2}\,\mathrm{erg\,cm}^{-3}\,,
  \end{split}
\end{equation}
where $R$ is the separation between the star and the pulsar (system separation). For simplicity, we assume that the production region is located close to the pulsar.

For a Gauss-strength magnetic field, VHE electrons generate synchrotron emission in the hard X-ray band. Binary pulsar systems were predicted, therefore, to be TeV sources, provided that the energy density of the stellar photon field is comparable to the expected energy density of the magnetic field \citep{1999APh....10...31K}. For an accurate calculation of the expected TeV flux level, one needs to account for a number of effects including the Klein-Nishina cutoff, IC scattering in the anisotropic regime, and gamma-gamma attenuation \citep{1999APh....10...31K}.

\begin{figure}
\centering
\includegraphics[width=1.0\columnwidth,center]{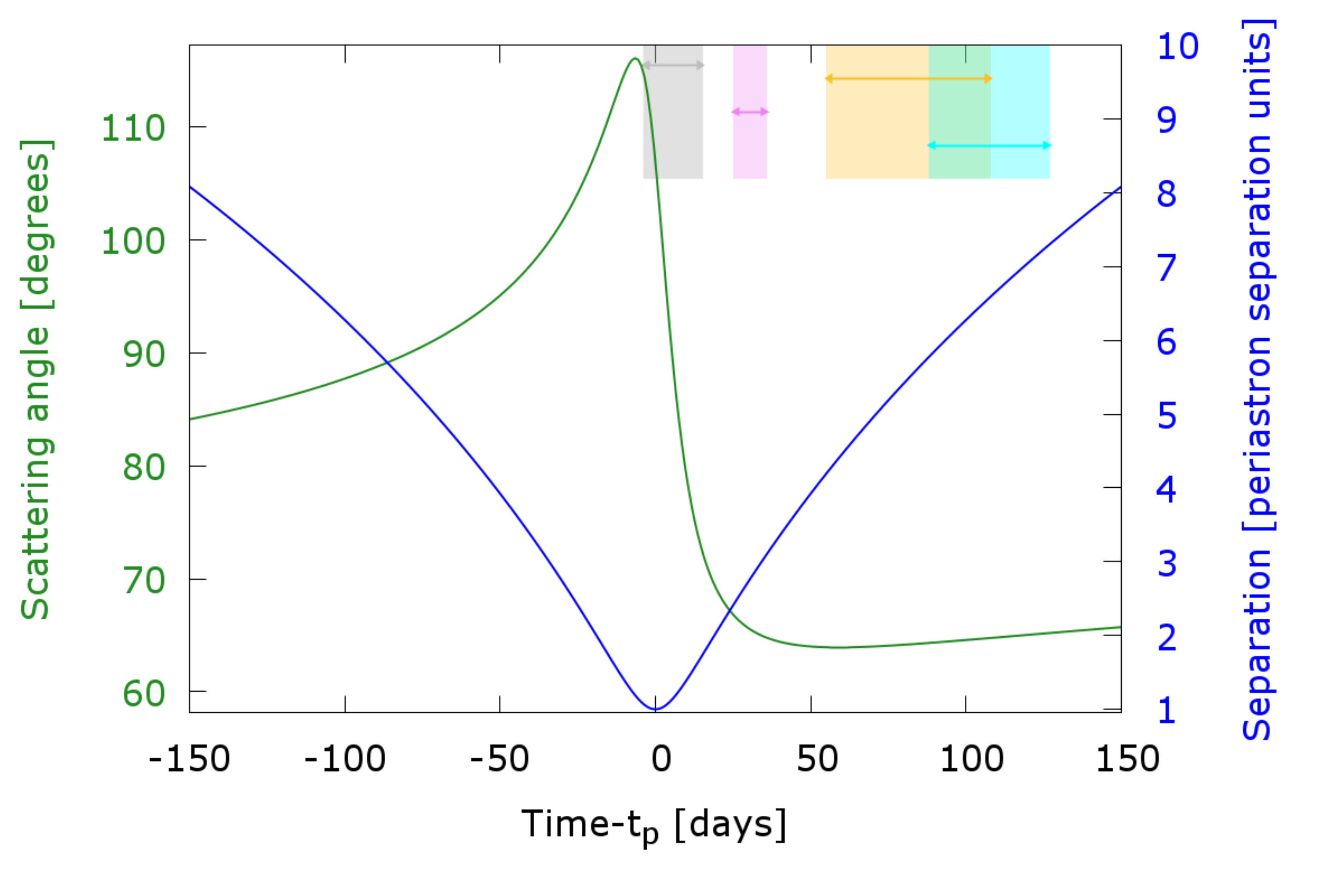}
    \caption{Comparison of the scattering angle of IC processes to separation of the two objects comprising \psrb. \textbf{Left axis, green curve:} The angle between the line-of-sight and the direction from the optical star at the pulsar's location, as a function of time to periastron passage. If the production region is close to the pulsar, this angle is equal to the scattering angle for IC processes. \textbf{Right axis, blue curve:} The ratio of the separation distance, $R$, to the periastron separation. The shaded regions in this plot represent the periods of the sub spectra and are defined as in Fig.~\ref{fig:lc} (see Tab.~\ref{table:spectra_details} for the full details of these sub-periods).}
    \label{fig:scattering_angle}
\end{figure}

Under the assumption of isotropic winds, the pulsar wind termination distance is proportional to the system separation distance, $R_{\textrm{ts}}\propto R$. Thus, the ratio of the photon to magnetic field energy density does not depend on the orbital phase, and one may expect quite similar X-ray and TeV light curves unless $\gamma$--$\gamma$ attenuation is significant \citep{2006A&A...451....9D,khangulyan_2007,sushch17,Sushch_2023}. However, one needs to take into account that some physical parameters can change their values depending on the orbital phase. For example, the magnetic field strength, which is expected to be proportional to the distance to the termination shock, may undergo a significant change with orbital phase. Consequently, this may induce a change of the cooling regime and/or of the synchrotron component \citep{Khangulyan_2012, dubus_and_cerutti_2013}.

Unless the stellar disc or locally generated fields provide significant targets for IC scattering, the temperature of the target photons does not vary with orbital phase. However, one needs to account for the change of the scattering angle. For an orbital inclination angle of $i\approx154^\circ$ \citep{millerjones_2018},
and for a production region in the orbital plane during the epochs of the \hess observations (see Fig.~\ref{fig:scattering_angle}), the IC scattering angle is approximately  $65^\circ$.
In this case, an emission of energy $1$ TeV may be generated by electrons with energy $E_{\mathrm{TeV}}\approx1.6$~TeV; here we adopt a photon field temperature of $T_*\approx3\times10^4$~K \citep{Negueruela_2011}, and use the approximation from \citet{2014ApJ...783..100K}.

Because of the eccentric orbit of the pulsar in the system, the system separation changes by a factor of four during the period relevant for the \hess observations.
This will therefore induce a proportional change of the magnetic field strength in the production region, meaning that the energy of the X-ray emitting electrons may change by a factor of $\approx2$.
For a typical X-ray spectrum slope of $1.5$ \citep[the average value obtained from \swift observations by][]{chernyakova_21}, X-ray emitting electrons have an $E^{-2}$ energy distribution and so, even in an idealised case of isotropic winds, the relationship between X-ray and TeV luminosity should depend on separation as:
\begin{equation}\label{eq:x-tev_cor}
  L_X\propto L_{\mathrm{TeV}} R^{1/2}\,.
\end{equation}

In Fig.~\ref{fig:Xray_TeV_linear}, the linear fit is mostly constrained by pairs of correlated X-ray and TeV runs occurring at $t>\tp+50$~days, i.e., when $R$ is large and changes more slowly with time.
For smaller separation distances, using Eq.~(\ref{eq:x-tev_cor}) one should expect that the linear fit overestimates the X-ray flux level.
However, from Fig.~\ref{fig:Xray_TeV_linear} one can see that for certain pairs the measured X-ray flux is significantly higher than the value given by the linear fit.
This could be considered as a hint of the wind interaction in a non-isotropic regime (e.g a Keplerian decretion disc, a non-isotropic pulsar wind, or changes in the scattering angle between relativistic electrons and soft photons).
Indeed, the points with high relative X-ray flux correspond to orbital phases close to $\tp+16$~days, where the pulsar may interact with the stellar disc.
Providing a significantly dense stellar disc, the pulsar wind will terminate significantly closer to the pulsar, enhancing the magnetic field strength without a proportional enhancement of the photon field \citep{khangulyan_2007,2012ApJ...750...70T,2014JHEAp...3...18S}.
This results in increasing X-ray flux and a softer X-ray spectrum during the disc crossing,  consistent with available observations~\citep[see e.g.,][]{chernyakova_14}.

Analysis of the X-ray -- TeV emission correlation reveals that there is a contribution to the X-ray flux that depends weakly on the TeV flux level (the contribution due to the constant term in the linear relationship).
This could indicate the presence of two or more zones that generate non-thermal emission. This is also supported by the absence of a strong correlation between the X-ray and radio emission.
Up to about 30 days after periastron, radio and X-ray emission show a very good correlation, but following this the X-ray flux starts to increase while the radio flux continues decreasing~\citep[see in][]{chernyakova_21}. 

This third X-ray peak, occuring 30-50 days after periastron \citep{chernyakova_21}, has not been reported in previous periastron passages. Although we lack TeV observations in 2021 during a larger fraction of this period, there is good evidence (from a significant TeV flux rise around $t_\mathbf{p}+30$ days, a good correlation with X-ray data in this period, and from the light curve template fitting discussed in Sect. 3.2) that there is a correspondence of the third X-ray peak in the 2021 TeV light curve. However, because of the gap in TeV coverage after $t_\mathbf{p}+35$ days, the time at which the maximum occurs in the TeV light curve is not well constrained and could be shifted with respect to the X-ray peak.

Summarising the multiwavelength data from \citet{chernyakova_21} during the period 30-60 days post-periastron, we see that the radio emission decreases, X-ray emission increases and then decreases, GeV emission stays in a low-emission state. From the \hess{} data, we see that the TeV emission increases and then decreases.
Immediately following this period, the GeV emission increases strongly with no corresponding increase in any other band.
Given the variation seen in emission profiles, it is difficult to reconcile all of these observational trends with a simple one-zone model for the post-periastron time-evolution of the non-thermal emission.
A multi-zone configuration can be produced by the complex geometry and dynamics of the interaction between the pulsar and stellar winds \citep{2008MNRAS.387...63B, 2012A&A...544A..59B, 2015A&A...581A..27D, 2021A&A...646A..91H}, and it appears likely that such models are required to explain the data.
The correlation of the TeV and X-ray light curves 30 days after periastron, suggests either that the electron population responsible for the third X-ray peak also emits in the TeV regime. Alternatively, given that the observed TeV light curve is compatible with previous periastron passages, it is possible that the X-ray emission accompanying the TeV peak was suppressed for some reason during previous periastron passages. 

The nature of the GeV flare, which is not accompanied by an increase in emission at any other waveband, remains puzzling. This scenario could potentially be connected to a complex evolution of the wind termination shock, i.e. strong confinement of the pulsar wind due to either the eccentricity of the orbit \citep{barkov_bosch-ramon_2016}, or the interaction with the circumstellar disc \citep{Khangulyan_2012} followed by a rapid expansion of the pulsar wind bubble later on.

\subsection{Spectral variability}
Another important finding in the \hess 2021 dataset is spectral variability of the VHE emission.
While in other GRLBs spectral variability is an established feature of the TeV emission \citep[most notably in LS~5039,][]{2005Sci...309..746A}, the previously reported VHE spectra of \psrb have a power-law shape with statistically indistinguishable photon indexes \citep{hess_psrb2020}.

In the context of GRLBs there are three major factors that cause changes of the VHE spectral slope: $\gamma$--$\gamma$ attenuation, anisotropic IC scattering, and changes in the distribution of emitting particles due to the orbital phase.
In the specific case of \psrb, $\gamma$--$\gamma$ attenuation might be relevant only at points close to the periastron passage which, most likely, has no significant impact during the orbital phases relevant for \hess observations in 2021.
Similarly, there is no significant change of the scattering angle during this period (see Fig.~\ref{fig:scattering_angle}).
With these aforementioned factors accounted for, the \hess spectral variation measurement implies a hardening of the electron distribution that could, for example, be caused by a change of the cooling regime.
If one assumes that the winds interact in an isotropic regime, then the rate of IC and synchrotron losses have a similar dependence ($\propto R^{-2}$) on the orbital phase.
On the other hand, the rate of adiabatic losses scale differently, $\propto R^{-1}$ \citep[see, e.g.,][]{2008IJMPD..17.1909K}.
Hence, one expects that at large system separations, the transition to an adiabatic loss-dominated cooling regime occurs at higher energies.

If this process indeed defines the hardening of the VHE spectrum, then one should also expect an analogous hardening during similar epochs prior to the periastron passage.
The stacked analysis of the \hess data collected in 2004, 2007, 2011, 2014, and 2017 indicates that VHE emission during the interval $\tp-109$~days to $\tp-47$~days has a photon index of $\Gamma=2.7\pm0.1_{\mathrm{stat}}\pm0.1_{\mathrm{sys}}$ \citep{hess_psrb2020} which is, in fact, significantly softer than the value obtained from ``symmetric'' orbital phases in 2021 (e.g., $\Gamma=2.42\pm0.1_{\mathrm{stat}}\pm0.1_{\mathrm{sys}}$ for the dataset C).
A complicating factor is that during the pre-periastron passage period the IC scattering angle is larger (see in Fig.~\ref{fig:scattering_angle}) and the resulting VHE spectrum is expected to be softer \citep[see, e.g.,][]{2008MNRAS.383..467K}.

In summary, it appears that the observed spectral change can be explained in the context of a hardening of the electron spectrum. This is, in turn, driven by changes in the scaling of cooling timescales as a result of varying orbital separation.
A detailed numerical model is required to quantitatively test the viability of such a scenario and is beyond the scope of this paper.
The possible important role of adiabatic losses supports the general conclusion that in binary pulsar systems, (magneto)hydrodynamic processes play an essential role for non-thermal radiation formation \citep{2008MNRAS.387...63B, 2012MNRAS.419.3426B, 2019MNRAS.490.3601B, 2012A&A...544A..59B}.
Hydrodynamic processes may also lead to the formation of several distinct production regions \citep{2013A&A...551A..17Z,2015A&A...581A..27D,2021A&A...646A..91H}, and the existence of these seem to be supported by observational evidence.
In particular we note the lack of a firm correlation between X-ray and radio emission \citep{chernyakova_21}, and the very different properties of the GeV and TeV emission detected from the system. A similar absence of correlation between GeV and TeV emission was seen during previous periastron passages~\citep{hess_psrb2020}.

\section{Conclusions}
\label{section:conclusions}

This work summarises the results from the \hess{} observations and analysis of the 2021 perisatron of the \psrb system in the VHE band. 
As displayed in Tab.~\ref{table:spectra_details}, our spectral studies reveal that the periastron-averaged spectrum can be described by a power-law model, with a spectral index of $\Gamma = 2.75 ~\pm~ 0.05 _{\text{stat}}$ $~\pm~ 0.1_{\text{sys}}$. This value is consistent with the average value reported in previous periastron passages ~\citep{hess_psrb2020}. We find that the fit has no preference to a power-law containing a cut-off component, with a lower limit on the cut-off energy of $E_{C}^{95\%} = 27.1$ TeV.
We also present, for the first time, evidence of spectral variability on a sub-orbital scale. A difference of $\Delta \Gamma = 0.56 ~\pm~ 0.18_{\text{stat}}$ $~\pm~ 0.10_{\text{sys}}$ (at greater than $95 \%$ c.i.) is seen between the spectral slopes of datasets B and D, see Tab.~\ref{table:spectra_details}. Since during the epochs corresponding to the datasets B and D, the $\gamma$--$\gamma$ absorption is negligible and the change of the IC scattering angle is small, the revealed hardening indicates on a change of the energy distribution of the emitting particles, which can be caused by a change of the cooling regime. 

The study of contemporaneous X-ray and TeV fluxes allowed the establishment of a linear correlation between the two energy bands. While the majority of the dataset is fitted relatively well by the applied linear fit (see in Fig. \ref{fig:lightcurve _correlated}), two data pairs show significantly higher X-ray flux levels. The two outliers correspond to orbital phases when the pulsar likely interacts with the disc, therefore the structure of the flow deviates considerably from an axially-symmetric configuration. During this period, it is expected that the pulsar wind terminates at a significantly smaller distance, thereby strongly enhancing the magnetic field.

Regarding the TeV data taken during the time period of the third X-ray peak, we argue that there is good evidence for a correspondence of this TeV data to the third X-ray peak, in the 2021 TeV light curve. However, the time of the maximum in the TeV light curve is not well constrained because of a lack of data 35-55 days post-periastron. Nevertheless, this feature is very interesting and requires further investigation.

The correlation obtained contains a significant constant term, which implies a presence of X-ray emitting electrons with no proportional TeV component. This supports the existence of a multiple emission zone geometry within the system. The evidence for a multi-zone setup can also be obtained from the uncorrelated radiation in the GeV and TeV energy bands, as well as from the absence of a strong X-ray -- radio correlation. The formation of a multi-zone setup can originate as a result of the complexity of the hydrodynamics within the pulsar and stellar wind interaction. The detection of spectral hardening at TeV energies after the 2021 periastron passage, together with the measured X-ray -- TeV correlation, provide new constraints that will contribute to building a consistent physical model for the multiwavelength emission from \psrb.

\section*{Acknowledgements}
The support of the Namibian authorities and of the University of Namibia in facilitating the construction and operation of H.E.S.S. is gratefully acknowledged, as is the support by the German Ministry for Education and Research (BMBF), the Max Planck Society, the Helmholtz Association, the French Ministry of Higher Education, Research and Innovation, the Centre National de la Recherche Scientifique (CNRS/IN2P3 and CNRS/INSU), the Commissariat à l’énergie atomique et aux énergies alternatives (CEA), the U.K. Science and Technology Facilities Council (STFC), the Irish Research Council (IRC) and the Science Foundation Ireland (SFI), the Polish Ministry of Education and Science, agreement no. 2021/WK/06, the South African Department of Science and Innovation and National Research Foundation, the University of Namibia, the National Commission on Research, Science \& Technology of Namibia (NCRST), the Austrian Federal Ministry of Education, Science and Research and the Austrian Science Fund (FWF), the Australian Research Council (ARC), the Japan Society for the Promotion of Science, the University of Amsterdam and the Science Committee of Armenia grant 21AG-1C085. We appreciate the excellent work of the technical support staff in Berlin, Zeuthen, Heidelberg, Palaiseau, Paris, Saclay, Tübingen and in Namibia in the construction and operation of the equipment. This work benefited from services provided by the H.E.S.S. Virtual Organisation, supported by the national resource providers of the EGI Federation.

\def\aj{AJ}%
\def\actaa{Acta Astron.}%
\def\araa{ARA\&A}%
\def\apj{ApJ}%
\def\apjl{ApJ}%
\def\apjs{ApJS}%
\def\ao{Appl.~Opt.}%
\def\apss{Ap\&SS}%
\def\aap{A\&A}%
\def\aapr{A\&A~Rev.}%
\def\aaps{A\&AS}%
\def\azh{AZh}%
\def\baas{BAAS}%
\def\bac{Bull. astr. Inst. Czechosl.}%
\def\caa{Chinese Astron. Astrophys.}%
\def\cjaa{Chinese J. Astron. Astrophys.}%
\def\icarus{Icarus}%
\def\jcap{J. Cosmology Astropart. Phys.}%
\def\jrasc{JRASC}%
\def\mnras{MNRAS}%
\def\memras{MmRAS}%
\def\na{New A}%
\def\nar{New A Rev.}%
\def\pasa{PASA}%
\def\pra{Phys.~Rev.~A}%
\def\prb{Phys.~Rev.~B}%
\def\prc{Phys.~Rev.~C}%
\def\prd{Phys.~Rev.~D}%
\def\pre{Phys.~Rev.~E}%
\def\prl{Phys.~Rev.~Lett.}%
\def\pasp{PASP}%
\def\pasj{PASJ}%
\def\qjras{QJRAS}%
\def\rmxaa{Rev. Mexicana Astron. Astrofis.}%
\def\skytel{S\&T}%
\def\solphys{Sol.~Phys.}%
\def\sovast{Soviet~Ast.}%
\def\ssr{Space~Sci.~Rev.}%
\def\zap{ZAp}%
\def\nat{Nature}%
\def\iaucirc{IAU~Circ.}%
\def\aplett{Astrophys.~Lett.}%
\def\apspr{Astrophys.~Space~Phys.~Res.}%
\def\bain{Bull.~Astron.~Inst.~Netherlands}%
\def\fcp{Fund.~Cosmic~Phys.}%
\def\gca{Geochim.~Cosmochim.~Acta}%
\def\grl{Geophys.~Res.~Lett.}%
\def\jcp{J.~Chem.~Phys.}%
\def\jgr{J.~Geophys.~Res.}%
\def\jqsrt{J.~Quant.~Spec.~Radiat.~Transf.}%
\def\memsai{Mem.~Soc.~Astron.~Italiana}%
\def\nphysa{Nucl.~Phys.~A}%
\def\physrep{Phys.~Rep.}%
\def\physscr{Phys.~Scr}%
\def\planss{Planet.~Space~Sci.}%
\def\procspie{Proc.~SPIE}%
\let\astap=\aap
\let\apjlett=\apjl
\let\apjsupp=\apjs
\let\applopt=\ao
\bibliographystyle{aa}
\setlength{\bibsep}{0pt plus 0.3ex} 
\bibliography{psrb}

\begin{thebibliography}{64}
\expandafter\ifx\csname natexlab\endcsname\relax\def\natexlab#1{#1}\fi

\bibitem[{{Abdo} {et~al.}(2011){Abdo}, {Ackermann}, {Ajello}, {Allafort}, {Ballet}, {Barbiellini}, {Bastieri}, {Bechtol}, {Bellazzini}, {Berenji}, {Blandford}, {Bonamente}, {Borgland}, {Bregeon}, {Brigida}, {Bruel}, {Buehler}, {Buson}, {Caliandro}, {Cameron}, {Camilo}, {Caraveo}, {Cecchi}, {Charles}, {Chaty}, {Chekhtman}, {Chernyakova}, {Cheung}, {Chiang}, {Ciprini}, {Claus}, {Cohen-Tanugi}, {Cominsky}, {Corbel}, {Cutini}, {D'Ammando}, {de Angelis}, {den Hartog}, {de Palma}, {Dermer}, {Digel}, {Silva}, {Dormody}, {Drell}, {Drlica-Wagner}, {Dubois}, {Dubus}, {Dumora}, {Enoto}, {Espinoza}, {Favuzzi}, {Fegan}, {Ferrara}, {Focke}, {Fortin}, {Fukazawa}, {Funk}, {Fusco}, {Gargano}, {Gasparrini}, {Gehrels}, {Germani}, {Giglietto}, {Giommi}, {Giordano}, {Giroletti}, {Glanzman}, {Godfrey}, {Grenier}, {Grondin}, {Grove}, {Grundstrom}, {Guiriec}, {Gwon}, {Hadasch}, {Harding}, {Hayashida}, {Hays}, {J{\'o}hannesson}, {Johnson}, {Johnson}, {Johnston}, {Kamae}, {Katagiri}, {Kataoka}, {Keith}, {Kerr}, {Kn{\"o}dlseder},
  {Kramer}, {Kuss}, {Lande}, {Lee}, {Lemoine-Goumard}, {Longo}, {Loparco}, {Lovellette}, {Lubrano}, {Manchester}, {Marelli}, {Mazziotta}, {Michelson}, {Mitthumsiri}, {Mizuno}, {Moiseev}, {Monte}, {Monzani}, {Morselli}, {Moskalenko}, {Murgia}, {Nakamori}, {Naumann-Godo}, {Neronov}, {Nolan}, {Norris}, {Noutsos}, {Nuss}, {Ohsugi}, {Okumura}, {Omodei}, {Orlando}, {Paneque}, {Parent}, {Pesce-Rollins}, {Pierbattista}, {Piron}, {Porter}, {Possenti}, {Rain{\`o}}, {Rando}, {Ray}, {Razzano}, {Razzaque}, {Reimer}, {Reimer}, {Reposeur}, {Ritz}, {Sadrozinski}, {Scargle}, {Sgr{\`o}}, {Shannon}, {Siskind}, {Smith}, {Spandre}, {Spinelli}, {Strickman}, {Suson}, {Takahashi}, {Tanaka}, {Thayer}, {Thayer}, {Thompson}, {Thorsett}, {Tibaldo}, {Tibolla}, {Torres}, {Tosti}, {Troja}, {Uchiyama}, {Usher}, {Vandenbroucke}, {Vasileiou}, {Vianello}, {Vitale}, {Waite}, {Wang}, {Winer}, {Wolff}, {Wood}, {Wood}, {Yang}, {Ziegler}, \& {Zimmer}}]{Abdo2011_b1259}
{Abdo}, A.~A., {Ackermann}, M., {Ajello}, M., {et~al.} 2011, \apjl, 736, L11

\bibitem[{{Abdo} {et~al.}(2009){Abdo}, {Ackermann}, {Ajello}, {Anderson}, {Atwood}, {Axelsson}, {Baldini}, {Ballet}, {Barbiellini}, {Baring}, {Bastieri}, {Baughman}, {Bechtol}, {Bellazzini}, {Berenji}, {Bignami}, {Blandford}, {Bloom}, {Bonamente}, {Borgland}, {Bregeon}, {Brez}, {Brigida}, {Bruel}, {Burnett}, {Caliandro}, {Cameron}, {Caraveo}, {Casandjian}, {Cecchi}, {{\c{C}}elik}, {Chekhtman}, {Cheung}, {Chiang}, {Ciprini}, {Claus}, {Cohen-Tanugi}, {Conrad}, {Cutini}, {Dermer}, {de Angelis}, {de Luca}, {de Palma}, {Digel}, {Dormody}, {do Couto e Silva}, {Drell}, {Dubois}, {Dumora}, {Farnier}, {Favuzzi}, {Fegan}, {Fukazawa}, {Funk}, {Fusco}, {Gargano}, {Gasparrini}, {Gehrels}, {Germani}, {Giebels}, {Giglietto}, {Giommi}, {Giordano}, {Glanzman}, {Godfrey}, {Grenier}, {Grondin}, {Grove}, {Guillemot}, {Guiriec}, {Gwon}, {Hanabata}, {Harding}, {Hayashida}, {Hays}, {Hughes}, {J{\'o}hannesson}, {Johnson}, {Johnson}, {Johnson}, {Kamae}, {Katagiri}, {Kataoka}, {Kawai}, {Kerr}, {Kn{\"o}dlseder}, {Kocian}, {Kuss},
  {Lande}, {Latronico}, {Lemoine-Goumard}, {Longo}, {Loparco}, {Lott}, {Lovellette}, {Lubrano}, {Madejski}, {Makeev}, {Marelli}, {Mazziotta}, {McConville}, {McEnery}, {Meurer}, {Michelson}, {Mitthumsiri}, {Mizuno}, {Monte}, {Monzani}, {Morselli}, {Moskalenko}, {Murgia}, {Nolan}, {Norris}, {Nuss}, {Ohsugi}, {Omodei}, {Orlando}, {Ormes}, {Paneque}, {Parent}, {Pelassa}, {Pepe}, {Pesce-Rollins}, {Pierbattista}, {Piron}, {Porter}, {Primack}, {Rain{\`o}}, {Rando}, {Ray}, {Razzano}, {Rea}, {Reimer}, {Reimer}, {Reposeur}, {Ritz}, {Rochester}, {Rodriguez}, {Romani}, {Ryde}, {Sadrozinski}, {Sanchez}, {Sander}, {Parkinson}, {Scargle}, {Sgr{\`o}}, {Siskind}, {Smith}, {Smith}, {Spandre}, {Spinelli}, {Starck}, {Strickman}, {Suson}, {Tajima}, {Takahashi}, {Takahashi}, {Tanaka}, {Thayer}, {Thompson}, {Tibaldo}, {Tibolla}, {Torres}, {Tosti}, {Tramacere}, {Uchiyama}, {Usher}, {Van Etten}, {Vasileiou}, {Vilchez}, {Vitale}, {Waite}, {Wang}, {Watters}, {Winer}, {Wolff}, {Wood}, {Ylinen}, {Ziegler}, \& {Fermi LAT
  Collaboration}}]{abdo_2009_J2032_detection}
{Abdo}, A.~A., {Ackermann}, M., {Ajello}, M., {et~al.} 2009, Science, 325, 840

\bibitem[{Abdo {et~al.}(2009)Abdo, Ackermann, Ajello, Atwood, Axelsson, Baldini, Ballet, Barbiellini, Baring, Bastieri, Bechtol, Bellazzini, Berenji, Bloom, Bonamente, Borgland, Bregeon, Brez, Brigida, Bruel, Burnett, Caliandro, Cameron, Caraveo, Casandjian, Cavazzuti, Cecchi, Çelik, Chekhtman, Cheung, Chiang, Ciprini, Claus, Cohen-Tanugi, Cominsky, Conrad, Cutini, de~Angelis, de~Palma, Bernardo, do~Couto~e Silva, Drell, Drlica-Wagner, Dubois, Dumora, Farnier, Favuzzi, Fegan, Finke, Focke, Fortin, Foschini, Frailis, Fukazawa, Funk, Fusco, Gargano, Gasparrini, Gehrels, Germani, Giavitto, Giebels, Giglietto, Giommi, Giordano, Glanzman, Godfrey, Grenier, Grondin, Grove, Guillemot, Guiriec, Hanabata, Hayashida, Hays, Horan, Hughes, Jackson, Jóhannesson, Johnson, Johnson, Johnson, Kamae, Katagiri, Kataoka, Kawai, Kerr, Knödlseder, Kocian, Kuss, Lande, Latronico, Lemoine-Goumard, Longo, Loparco, Lott, Lovellette, Lubrano, Madejski, Makeev, Mazziotta, McConville, McEnery, Meurer, Michelson, Mitthumsiri, Mizuno,
  Moiseev, Monte, Monzani, Morselli, Moskalenko, Murgia, Nolan, Norris, Nuss, Ohsugi, Omodei, Orlando, Ormes, Ozaki, Paneque, Panetta, Parent, Pelassa, Pepe, Pesce-Rollins, Piron, Porter, Rainò, Rando, Razzano, Reimer, Reimer, Reposeur, Reyes, Ritz, Rochester, Rodriguez, Roth, Ryde, Sadrozinski, Sanchez, Sander, Parkinson, Scargle, Schalk, Sellerholm, Sgrò, Shaw, Siskind, Smith, Smith, Spandre, Spinelli, Strickman, Suson, Tajima, Takahashi, Takahashi, Tanaka, Tanaka, Thayer, Thayer, Thompson, Tibaldo, Torres, Tosti, Tramacere, Uchiyama, Usher, Vasileiou, Vilchez, Vitale, Waite, Wang, Winer, Wood, Ylinen, \& Ziegler}]{Abdo_2009}
Abdo, A.~A., Ackermann, M., Ajello, M., {et~al.} 2009, The Astrophysical Journal, 707, 1310

\bibitem[{Acero {et~al.}(2013)Acero, Ackermann, Ajello, Allafort, Baldini, Ballet, Barbiellini, Bastieri, Bechtol, Bellazzini, Blandford, Bloom, Bonamente, Bottacini, Brandt, Bregeon, Brigida, Bruel, Buehler, Buson, Caliandro, Cameron, Caraveo, Cecchi, Charles, Chaves, Chekhtman, Chiang, Chiaro, Ciprini, Claus, Cohen-Tanugi, Conrad, Cutini, Dalton, D{\textquotesingle}Ammando, de~Palma, Dermer, Venere, do~Couto~e Silva, Drell, Drlica-Wagner, Falletti, Favuzzi, Fegan, Ferrara, Focke, Franckowiak, Fukazawa, Funk, Fusco, Gargano, Gasparrini, Giglietto, Giordano, Giroletti, Glanzman, Godfrey, Gr{\'{e} }goire, Grenier, Grondin, Grove, Guiriec, Hadasch, Hanabata, Harding, Hayashida, Hayashi, Hays, Hewitt, Hill, Horan, Hou, Hughes, Inoue, Jackson, Jogler, J{\'{o}}hannesson, Johnson, Kamae, Kawano, Kerr, Knödlseder, Kuss, Lande, Larsson, Latronico, Lemoine-Goumard, Longo, Loparco, Lovellette, Lubrano, Marelli, Massaro, Mayer, Mazziotta, McEnery, Mehault, Michelson, Mitthumsiri, Mizuno, Monte, Monzani, Morselli,
  Moskalenko, Murgia, Nakamori, Nemmen, Nuss, Ohsugi, Okumura, Orienti, Orlando, Ormes, Paneque, Panetta, Perkins, Pesce-Rollins, Piron, Pivato, Porter, Rain{\`{o}}, Rando, Razzano, Reimer, Reimer, Reposeur, Ritz, Roth, Rousseau, Parkinson, Schulz, Sgr{\`{o}}, Siskind, Smith, Spandre, Spinelli, Suson, Takahashi, Takeuchi, Thayer, Thayer, Thompson, Tibaldo, Tibolla, Tinivella, Torres, Tosti, Troja, Uchiyama, Vandenbroucke, Vasileiou, Vianello, Vitale, Werner, Winer, Wood, \& Yang}]{Acero_2013}
Acero, F., Ackermann, M., Ajello, M., {et~al.} 2013, The Astrophysical Journal, 773, 77

\bibitem[{{Aharonian} {et~al.}(2009){Aharonian}, {Akhperjanian}, {Anton}, {Barres de Almeida}, {Bazer-Bachi}, {Becherini}, {Behera}, {Bernl{\"o}hr}, {Bochow}, {Boisson}, {Bolmont}, {Borrel}, {Brucker}, {Brun}, {Brun}, {B{\"u}hler}, {Bulik}, {B{\"u}sching}, {Boutelier}, {Chadwick}, {Charbonnier}, {Chaves}, {Cheesebrough}, {Chounet}, {Clapson}, {Coignet}, {Dalton}, {Daniel}, {Davids}, {Degrange}, {Deil}, {Dickinson}, {Djannati-Ata{\"\i}}, {Domainko}, {O'C. Drury}, {Dubois}, {Dubus}, {Dyks}, {Dyrda}, {Egberts}, {Emmanoulopoulos}, {Espigat}, {Farnier}, {Feinstein}, {Fiasson}, {F{\"o}rster}, {Fontaine}, {F{\"u}{\ss}ling}, {Gabici}, {Gallant}, {G{\'e}rard}, {Gerbig}, {Giebels}, {Glicenstein}, {Gl{\"u}ck}, {Goret}, {G{\"o}ring}, {Hauser}, {Hauser}, {Heinz}, {Heinzelmann}, {Henri}, {Hermann}, {Hinton}, {Hoffmann}, {Hofmann}, {Holleran}, {Hoppe}, {Horns}, {Jacholkowska}, {de Jager}, {Jahn}, {Jung}, {Katarzy{\'n}ski}, {Katz}, {Kaufmann}, {Kerschhaggl}, {Khangulyan}, {Kh{\'e}lifi}, {Keogh}, {Klochkov}, {Klu{\'z}niak},
  {Kneiske}, {Komin}, {Kosack}, {Kossakowski}, {Lamanna}, {Lenain}, {Lohse}, {Marandon}, {Martineau-Huynh}, {Marcowith}, {Masbou}, {Maurin}, {McComb}, {Medina}, {Moderski}, {Moulin}, {Naumann-Godo}, {de Naurois}, {Nedbal}, {Nekrassov}, {Nicholas}, {Niemiec}, {Nolan}, {Ohm}, {Olive}, {de O{\~n}a Wilhelmi}, {Orford}, {Ostrowski}, {Panter}, {Paz Arribas}, {Pedaletti}, {Pelletier}, {Petrucci}, {Pita}, {P{\"u}hlhofer}, {Punch}, {Quirrenbach}, {Raubenheimer}, {Raue}, {Rayner}, {Renaud}, {Rieger}, {Ripken}, {Rob}, {Rosier-Lees}, {Rowell}, {Rudak}, {Rulten}, {Ruppel}, {Sahakian}, {Santangelo}, {Schlickeiser}, {Sch{\"o}ck}, {Schwanke}, {Schwarzburg}, {Schwemmer}, {Shalchi}, {Sikora}, {Skilton}, {Sol}, {Spangler}, {Stawarz}, {Steenkamp}, {Stegmann}, {Stinzing}, {Superina}, {Szostek}, {Tam}, {Tavernet}, {Terrier}, {Tibolla}, {Tluczykont}, {van Eldik}, {Vasileiadis}, {Venter}, {Venter}, {Vialle}, {Vincent}, {Vivier}, {V{\"o}lk}, {Volpe}, {Wagner}, {Ward}, {Zdziarski}, \& {Zech}}]{psrb2007}
{Aharonian}, F., {Akhperjanian}, A.~G., {Anton}, G., {et~al.} 2009, \aap, 507, 389

\bibitem[{{Aharonian} {et~al.}(2005{\natexlab{a}}){Aharonian}, {Akhperjanian}, {Aye}, {Bazer-Bachi}, {Beilicke}, {Benbow}, {Berge}, {Berghaus}, {Bernl{\"o}hr}, {Boisson}, {Bolz}, {Borrel}, {Braun}, {Breitling}, {Brown}, {Gordo}, {Chadwick}, {Chounet}, {Cornils}, {Costamante}, {Degrange}, {Dickinson}, {Djannati-Ata{\"\i}}, {Drury}, {Dubus}, {Emmanoulopoulos}, {Espigat}, {Feinstein}, {Fleury}, {Fontaine}, {Fuchs}, {Funk}, {Gallant}, {Giebels}, {Gillessen}, {Glicenstein}, {Goret}, {Hadjichristidis}, {Hauser}, {Heinzelmann}, {Henri}, {Hermann}, {Hinton}, {Hofmann}, {Holleran}, {Horns}, {Jacholkowska}, {de Jager}, {Kh{\'e}lifi}, {Komin}, {Konopelko}, {Latham}, {Le Gallou}, {Lemi{\`e}re}, {Lemoine-Goumard}, {Leroy}, {Lohse}, {Marcowith}, {Martin}, {Martineau-Huynh}, {Masterson}, {McComb}, {de Naurois}, {Nolan}, {Noutsos}, {Orford}, {Osborne}, {Ouchrif}, {Panter}, {Pelletier}, {Pita}, {P{\"u}hlhofer}, {Punch}, {Raubenheimer}, {Raue}, {Raux}, {Rayner}, {Reimer}, {Reimer}, {Ripken}, {Rob}, {Rolland}, {Rowell},
  {Sahakian}, {Saug{\'e}}, {Schlenker}, {Schlickeiser}, {Schuster}, {Schwanke}, {Siewert}, {Sol}, {Spangler}, {Steenkamp}, {Stegmann}, {Tavernet}, {Terrier}, {Th{\'e}oret}, {Tluczykont}, {Vasileiadis}, {Venter}, {Vincent}, {V{\"o}lk}, \& {Wagner}}]{2005Sci...309..746A}
{Aharonian}, F., {Akhperjanian}, A.~G., {Aye}, K.~M., {et~al.} 2005{\natexlab{a}}, Science, 309, 746

\bibitem[{{Aharonian} {et~al.}(2005{\natexlab{b}}){Aharonian}, {Akhperjanian}, {Aye}, {Bazer-Bachi}, {Beilicke}, {Benbow}, {Berge}, {Berghaus}, {Bernl{\"o}hr}, {Boisson}, {Bolz}, {Braun}, {Breitling}, {Brown}, {Bussons Gordo}, {Chadwick}, {Chounet}, {Cornils}, {Costamante}, {Degrange}, {Djannati-Ata{\"i}}, {O'C.~Drury}, {Dubus}, {Emmanoulopoulos}, {Espigat}, {Feinstein}, {Fleury}, {Fontaine}, {Fuchs}, {Funk}, {Gallant}, {Giebels}, {Gillessen}, {Glicenstein}, {Goret}, {Hadjichristidis}, {Hauser}, {Heinzelmann}, {Henri}, {Hermann}, {Hinton}, {Hofmann}, {Holleran}, {Horns}, {de Jager}, {Johnston}, {Kh{\'e}lifi}, {Kirk}, {Komin}, {Konopelko}, {Latham}, {Le Gallou}, {Lemi{\`e}re}, {Lemoine-Goumard}, {Leroy}, {Martineau-Huynh}, {Lohse}, {Marcowith}, {Masterson}, {McComb}, {de Naurois}, {Nolan}, {Noutsos}, {Orford}, {Osborne}, {Ouchrif}, {Panter}, {Pelletier}, {Pita}, {P{\"u}hlhofer}, {Punch}, {Raubenheimer}, {Raue}, {Raux}, {Rayner}, {Redondo}, {Reimer}, {Reimer}, {Ripken}, {Rob}, {Rolland}, {Rowell}, {Sahakian},
  {Saug{\'e}}, {Schlenker}, {Schlickeiser}, {Schuster}, {Schwanke}, {Siewert}, {Skj{\ae}raasen}, {Sol}, {Steenkamp}, {Stegmann}, {Tavernet}, {Terrier}, {Th{\'e}oret}, {Tluczykont}, {Vasileiadis}, {Venter}, {Vincent}, {V{\"o}lk}, \& {Wagner}}]{psrb_hess_discovery}
{Aharonian}, F., {Akhperjanian}, A.~G., {Aye}, K.-M., {et~al.} 2005{\natexlab{b}}, \aap, 442, 1

\bibitem[{{Aharonian} {et~al.}(2005{\natexlab{c}}){Aharonian}, {Akhperjanian}, {Aye}, {Bazer-Bachi}, {Beilicke}, {Benbow}, {Berge}, {Berghaus}, {Bernl{\"o}hr}, {Boisson}, {Bolz}, {Braun}, {Breitling}, {Brown}, {Bussons Gordo}, {Chadwick}, {Chounet}, {Cornils}, {Costamante}, {Degrange}, {Djannati-Ata{\"\i}}, {O'C. Drury}, {Dubus}, {Emmanoulopoulos}, {Espigat}, {Feinstein}, {Fleury}, {Fontaine}, {Fuchs}, {Funk}, {Gallant}, {Giebels}, {Gillessen}, {Glicenstein}, {Goret}, {Hadjichristidis}, {Hauser}, {Heinzelmann}, {Henri}, {Hermann}, {Hinton}, {Hofmann}, {Holleran}, {Horns}, {de Jager}, {Johnston}, {Kh{\'e}lifi}, {Kirk}, {Komin}, {Konopelko}, {Latham}, {Le Gallou}, {Lemi{\`e}re}, {Lemoine-Goumard}, {Leroy}, {Martineau-Huynh}, {Lohse}, {Marcowith}, {Masterson}, {McComb}, {de Naurois}, {Nolan}, {Noutsos}, {Orford}, {Osborne}, {Ouchrif}, {Panter}, {Pelletier}, {Pita}, {P{\"u}hlhofer}, {Punch}, {Raubenheimer}, {Raue}, {Raux}, {Rayner}, {Redondo}, {Reimer}, {Reimer}, {Ripken}, {Rob}, {Rolland}, {Rowell}, {Sahakian},
  {Saug{\'e}}, {Schlenker}, {Schlickeiser}, {Schuster}, {Schwanke}, {Siewert}, {Skj{\ae}raasen}, {Sol}, {Steenkamp}, {Stegmann}, {Tavernet}, {Terrier}, {Th{\'e}oret}, {Tluczykont}, {Vasileiadis}, {Venter}, {Vincent}, {V{\"o}lk}, \& {Wagner}}]{psrb2004}
{Aharonian}, F., {Akhperjanian}, A.~G., {Aye}, K.~M., {et~al.} 2005{\natexlab{c}}, \aap, 442, 1

\bibitem[{{Aharonian} {et~al.}(2006){Aharonian}, {Akhperjanian}, {Bazer-Bachi}, {Beilicke}, {Benbow}, {Berge}, {Bernl{\"o}hr}, {Boisson}, {Bolz}, {Borrel}, {Braun}, {Breitling}, {Brown}, {Chadwick}, {Chounet}, {Cornils}, {Costamante}, {Degrange}, {Dickinson}, {Djannati-Ata{\"\i}}, {Drury}, {Dubus}, {Emmanoulopoulos}, {Espigat}, {Feinstein}, {Fontaine}, {Fuchs}, {Funk}, {Gallant}, {Giebels}, {Gillessen}, {Glicenstein}, {Goret}, {Hadjichristidis}, {Hauser}, {Heinzelmann}, {Henri}, {Hermann}, {Hinton}, {Hofmann}, {Holleran}, {Horns}, {Jacholkowska}, {de Jager}, {Kh{\'e}lifi}, {Komin}, {Konopelko}, {Latham}, {Le Gallou}, {Lemi{\`e}re}, {Lemoine-Goumard}, {Leroy}, {Lohse}, {Martin}, {Martineau-Huynh}, {Marcowith}, {Masterson}, {McComb}, {de Naurois}, {Nolan}, {Noutsos}, {Orford}, {Osborne}, {Ouchrif}, {Panter}, {Pelletier}, {Pita}, {P{\"u}hlhofer}, {Punch}, {Raubenheimer}, {Raue}, {Raux}, {Rayner}, {Reimer}, {Reimer}, {Ripken}, {Rob}, {Rolland}, {Rowell}, {Sahakian}, {Saug{\'e}}, {Schlenker}, {Schlickeiser},
  {Schuster}, {Schwanke}, {Siewert}, {Sol}, {Spangler}, {Steenkamp}, {Stegmann}, {Tavernet}, {Terrier}, {Th{\'e}oret}, {Tluczykont}, {Vasileiadis}, {Venter}, {Vincent}, {V{\"o}lk}, \& {Wagner}}]{aharonian_06}
{Aharonian}, F., {Akhperjanian}, A.~G., {Bazer-Bachi}, A.~R., {et~al.} 2006, \apj, 636, 777

\bibitem[{{Barkov} \& {Bosch-Ramon}(2016)}]{barkov_bosch-ramon_2016}
{Barkov}, M.~V. \& {Bosch-Ramon}, V. 2016, \mnras, 456, L64

\bibitem[{Bausch(2013)}]{bausch_2013}
Bausch, J. 2013, Journal of Physics A: Mathematical and Theoretical, 46, 505202

\bibitem[{{Berge} {et~al.}(2007){Berge}, {Funk}, \& {Hinton}}]{Berge_07}
{Berge}, D., {Funk}, S., \& {Hinton}, J. 2007, \aap, 466, 1219

\bibitem[{{Bogovalov} {et~al.}(2019){Bogovalov}, {Khangulyan}, {Koldoba}, {Ustyugova}, \& {Aharonian}}]{2019MNRAS.490.3601B}
{Bogovalov}, S.~V., {Khangulyan}, D., {Koldoba}, A., {Ustyugova}, G.~V., \& {Aharonian}, F. 2019, \mnras, 490, 3601

\bibitem[{{Bogovalov} {et~al.}(2012){Bogovalov}, {Khangulyan}, {Koldoba}, {Ustyugova}, \& {Aharonian}}]{2012MNRAS.419.3426B}
{Bogovalov}, S.~V., {Khangulyan}, D., {Koldoba}, A.~V., {Ustyugova}, G.~V., \& {Aharonian}, F.~A. 2012, \mnras, 419, 3426

\bibitem[{{Bogovalov} {et~al.}(2008){Bogovalov}, {Khangulyan}, {Koldoba}, {Ustyugova}, \& {Aharonian}}]{2008MNRAS.387...63B}
{Bogovalov}, S.~V., {Khangulyan}, D.~V., {Koldoba}, A.~V., {Ustyugova}, G.~V., \& {Aharonian}, F.~A. 2008, \mnras, 387, 63

\bibitem[{{Bosch-Ramon} {et~al.}(2012){Bosch-Ramon}, {Barkov}, {Khangulyan}, \& {Perucho}}]{2012A&A...544A..59B}
{Bosch-Ramon}, V., {Barkov}, M.~V., {Khangulyan}, D., \& {Perucho}, M. 2012, \aap, 544, A59

\bibitem[{{Caliandro} {et~al.}(2015){Caliandro}, {Cheung}, {Li}, {Scargle}, {Torres}, {Wood}, \& {Chernyakova}}]{caliandro_2015}
{Caliandro}, G.~A., {Cheung}, C.~C., {Li}, J., {et~al.} 2015, \apj, 811, 68

\bibitem[{{Chang} {et~al.}(2021){Chang}, {Zhang}, {Chen}, {Ji}, {Kong}, \& {Wang}}]{chang_2021}
{Chang}, Z., {Zhang}, S., {Chen}, Y.-P., {et~al.} 2021, Universe, 7, 472

\bibitem[{{Chernyakova} {et~al.}(2014){Chernyakova}, {Abdo}, {Neronov}, {McSwain}, {Mold{\'o}n}, {Rib{\'o}}, {Paredes}, {Sushch}, {de Naurois}, {Schwanke}, {Uchiyama}, {Wood}, {Johnston}, {Chaty}, {Coleiro}, {Malyshev}, \& {Babyk}}]{chernyakova_14}
{Chernyakova}, M., {Abdo}, A.~A., {Neronov}, A., {et~al.} 2014, \mnras, 439, 432

\bibitem[{{Chernyakova} {et~al.}(2019){Chernyakova}, {Malyshev}, {Paizis}, {La Palombara}, {Balbo}, {Walter}, {Hnatyk}, {van Soelen}, {Romano}, {Munar-Adrover}, {Vovk}, {Piano}, {Capitanio}, {Falceta-Gon{\c{c}}alves}, {Landoni}, {Luque-Escamilla}, {Mart{\'\i}}, {Paredes}, {Rib{\'o}}, {Safi-Harb}, {Saha}, {Sidoli}, \& {Vercellone}}]{grlbcta19}
{Chernyakova}, M., {Malyshev}, D., {Paizis}, A., {et~al.} 2019, \aap, 631, A177

\bibitem[{{Chernyakova} {et~al.}(2021){Chernyakova}, {Malyshev}, {van Soelen}, {O'Sullivan}, {Sobey}, {Tsygankov}, {Mc Keague}, {Green}, {Kirwan}, {Santangelo}, {P{\"u}hlhofer}, \& {Monageng}}]{chernyakova_21}
{Chernyakova}, M., {Malyshev}, D., {van Soelen}, B., {et~al.} 2021, Universe, 7, 242

\bibitem[{{Chernyakova} {et~al.}(2015){Chernyakova}, {Neronov}, {van Soelen}, {Callanan}, {O'Shaughnessy}, {Babyk}, {Tsygankov}, {Vovk}, {Krivonos}, {Tomsick}, {Malyshev}, {Li}, {Wood}, {Torres}, {Zhang}, {Kretschmar}, {McSwain}, {Buckley}, \& {Koen}}]{chernyakova15}
{Chernyakova}, M., {Neronov}, A., {van Soelen}, B., {et~al.} 2015, \mnras, 454, 1358

\bibitem[{{Cominsky} {et~al.}(1994){Cominsky}, {Roberts}, \& {Johnston}}]{cominsky94}
{Cominsky}, L., {Roberts}, M., \& {Johnston}, S. 1994, \apj, 427, 978

\bibitem[{de~Naurois \& Rolland(2009)}]{de_Naurois_2009}
de~Naurois, M. \& Rolland, L. 2009, Astroparticle Physics, 32, 231

\bibitem[{{Dubus}(2006)}]{2006A&A...451....9D}
{Dubus}, G. 2006, \aap, 451, 9

\bibitem[{{Dubus}(2013)}]{dubus13}
{Dubus}, G. 2013, \aapr, 21, 64

\bibitem[{{Dubus} \& {Cerutti}(2013)}]{dubus_and_cerutti_2013}
{Dubus}, G. \& {Cerutti}, B. 2013, \aap, 557, A127

\bibitem[{{Dubus} {et~al.}(2015){Dubus}, {Lamberts}, \& {Fromang}}]{2015A&A...581A..27D}
{Dubus}, G., {Lamberts}, A., \& {Fromang}, S. 2015, \aap, 581, A27

\bibitem[{{Gaia Collaboration} {et~al.}(2018){Gaia Collaboration}, {Brown}, {Vallenari}, {Prusti}, {de Bruijne}, {Babusiaux}, {Bailer-Jones}, {Biermann}, {Evans}, {Eyer}, {Jansen}, {Jordi}, {Klioner}, {Lammers}, {Lindegren}, {Luri}, {Mignard}, {Panem}, {Pourbaix}, {Randich}, {Sartoretti}, {Siddiqui}, {Soubiran}, {van Leeuwen}, {Walton}, {Arenou}, {Bastian}, {Cropper}, {Drimmel}, {Katz}, {Lattanzi}, {Bakker}, {Cacciari}, {Casta{\~n}eda}, {Chaoul}, {Cheek}, {De Angeli}, {Fabricius}, {Guerra}, {Holl}, {Masana}, {Messineo}, {Mowlavi}, {Nienartowicz}, {Panuzzo}, {Portell}, {Riello}, {Seabroke}, {Tanga}, {Th{\'e}venin}, {Gracia-Abril}, {Comoretto}, {Garcia-Reinaldos}, {Teyssier}, {Altmann}, {Andrae}, {Audard}, {Bellas-Velidis}, {Benson}, {Berthier}, {Blomme}, {Burgess}, {Busso}, {Carry}, {Cellino}, {Clementini}, {Clotet}, {Creevey}, {Davidson}, {De Ridder}, {Delchambre}, {Dell'Oro}, {Ducourant}, {Fern{\'a}ndez-Hern{\'a}ndez}, {Fouesneau}, {Fr{\'e}mat}, {Galluccio}, {Garc{\'\i}a-Torres},
  {Gonz{\'a}lez-N{\'u}{\~n}ez}, {Gonz{\'a}lez-Vidal}, {Gosset}, {Guy}, {Halbwachs}, {Hambly}, {Harrison}, {Hern{\'a}ndez}, {Hestroffer}, {Hodgkin}, {Hutton}, {Jasniewicz}, {Jean-Antoine-Piccolo}, {Jordan}, {Korn}, {Krone-Martins}, {Lanzafame}, {Lebzelter}, {L{\"o}ffler}, {Manteiga}, {Marrese}, {Mart{\'\i}n-Fleitas}, {Moitinho}, {Mora}, {Muinonen}, {Osinde}, {Pancino}, {Pauwels}, {Petit}, {Recio-Blanco}, {Richards}, {Rimoldini}, {Robin}, {Sarro}, {Siopis}, {Smith}, {Sozzetti}, {S{\"u}veges}, {Torra}, {van Reeven}, {Abbas}, {Abreu Aramburu}, {Accart}, {Aerts}, {Altavilla}, {{\'A}lvarez}, {Alvarez}, {Alves}, {Anderson}, {Andrei}, {Anglada Varela}, {Antiche}, {Antoja}, {Arcay}, {Astraatmadja}, {Bach}, {Baker}, {Balaguer-N{\'u}{\~n}ez}, {Balm}, {Barache}, {Barata}, {Barbato}, {Barblan}, {Barklem}, {Barrado}, {Barros}, {Barstow}, {Bartholom{\'e} Mu{\~n}oz}, {Bassilana}, {Becciani}, {Bellazzini}, {Berihuete}, {Bertone}, {Bianchi}, {Bienaym{\'e}}, {Blanco-Cuaresma}, {Boch}, {Boeche}, {Bombrun}, {Borrachero},
  {Bossini}, {Bouquillon}, {Bourda}, {Bragaglia}, {Bramante}, {Breddels}, {Bressan}, {Brouillet}, {Br{\"u}semeister}, {Brugaletta}, {Bucciarelli}, {Burlacu}, {Busonero}, {Butkevich}, {Buzzi}, {Caffau}, {Cancelliere}, {Cannizzaro}, {Cantat-Gaudin}, {Carballo}, {Carlucci}, {Carrasco}, {Casamiquela}, {Castellani}, {Castro-Ginard}, {Charlot}, {Chemin}, {Chiavassa}, {Cocozza}, {Costigan}, {Cowell}, {Crifo}, {Crosta}, {Crowley}, {Cuypers}, {Dafonte}, {Damerdji}, {Dapergolas}, {David}, {David}, {de Laverny}, {De Luise}, {De March}, {de Martino}, {de Souza}, {de Torres}, {Debosscher}, {del Pozo}, {Delbo}, {Delgado}, {Delgado}, {Di Matteo}, {Diakite}, {Diener}, {Distefano}, {Dolding}, {Drazinos}, {Dur{\'a}n}, {Edvardsson}, {Enke}, {Eriksson}, {Esquej}, {Eynard Bontemps}, {Fabre}, {Fabrizio}, {Faigler}, {Falc{\~a}o}, {Farr{\`a}s Casas}, {Federici}, {Fedorets}, {Fernique}, {Figueras}, {Filippi}, {Findeisen}, {Fonti}, {Fraile}, {Fraser}, {Fr{\'e}zouls}, {Gai}, {Galleti}, {Garabato}, {Garc{\'\i}a-Sedano}, {Garofalo},
  {Garralda}, {Gavel}, {Gavras}, {Gerssen}, {Geyer}, {Giacobbe}, {Gilmore}, {Girona}, {Giuffrida}, {Glass}, {Gomes}, {Granvik}, {Gueguen}, {Guerrier}, {Guiraud}, {Guti{\'e}rrez-S{\'a}nchez}, {Haigron}, {Hatzidimitriou}, {Hauser}, {Haywood}, {Heiter}, {Helmi}, {Heu}, {Hilger}, {Hobbs}, {Hofmann}, {Holland}, {Huckle}, {Hypki}, {Icardi}, {Jan{\ss}en}, {Jevardat de Fombelle}, {Jonker}, {Juh{\'a}sz}, {Julbe}, {Karampelas}, {Kewley}, {Klar}, {Kochoska}, {Kohley}, {Kolenberg}, {Kontizas}, {Kontizas}, {Koposov}, {Kordopatis}, {Kostrzewa-Rutkowska}, {Koubsky}, {Lambert}, {Lanza}, {Lasne}, {Lavigne}, {Le Fustec}, {Le Poncin-Lafitte}, {Lebreton}, {Leccia}, {Leclerc}, {Lecoeur-Taibi}, {Lenhardt}, {Leroux}, {Liao}, {Licata}, {Lindstr{\o}m}, {Lister}, {Livanou}, {Lobel}, {L{\'o}pez}, {Managau}, {Mann}, {Mantelet}, {Marchal}, {Marchant}, {Marconi}, {Marinoni}, {Marschalk{\'o}}, {Marshall}, {Martino}, {Marton}, {Mary}, {Massari}, {Matijevi{\v{c}}}, {Mazeh}, {McMillan}, {Messina}, {Michalik}, {Millar}, {Molina}, {Molinaro},
  {Moln{\'a}r}, {Montegriffo}, {Mor}, {Morbidelli}, {Morel}, {Morris}, {Mulone}, {Muraveva}, {Musella}, {Nelemans}, {Nicastro}, {Noval}, {O'Mullane}, {Ord{\'e}novic}, {Ord{\'o}{\~n}ez-Blanco}, {Osborne}, {Pagani}, {Pagano}, {Pailler}, {Palacin}, {Palaversa}, {Panahi}, {Pawlak}, {Piersimoni}, {Pineau}, {Plachy}, {Plum}, {Poggio}, {Poujoulet}, {Pr{\v{s}}a}, {Pulone}, {Racero}, {Ragaini}, {Rambaux}, {Ramos-Lerate}, {Regibo}, {Reyl{\'e}}, {Riclet}, {Ripepi}, {Riva}, {Rivard}, {Rixon}, {Roegiers}, {Roelens}, {Romero-G{\'o}mez}, {Rowell}, {Royer}, {Ruiz-Dern}, {Sadowski}, {Sagrist{\`a} Sell{\'e}s}, {Sahlmann}, {Salgado}, {Salguero}, {Sanna}, {Santana-Ros}, {Sarasso}, {Savietto}, {Schultheis}, {Sciacca}, {Segol}, {Segovia}, {S{\'e}gransan}, {Shih}, {Siltala}, {Silva}, {Smart}, {Smith}, {Solano}, {Solitro}, {Sordo}, {Soria Nieto}, {Souchay}, {Spagna}, {Spoto}, {Stampa}, {Steele}, {Steidelm{\"u}ller}, {Stephenson}, {Stoev}, {Suess}, {Surdej}, {Szabados}, {Szegedi-Elek}, {Tapiador}, {Taris}, {Tauran}, {Taylor},
  {Teixeira}, {Terrett}, {Teyssandier}, {Thuillot}, {Titarenko}, {Torra Clotet}, {Turon}, {Ulla}, {Utrilla}, {Uzzi}, {Vaillant}, {Valentini}, {Valette}, {van Elteren}, {Van Hemelryck}, {van Leeuwen}, {Vaschetto}, {Vecchiato}, {Veljanoski}, {Viala}, {Vicente}, {Vogt}, {von Essen}, {Voss}, {Votruba}, {Voutsinas}, {Walmsley}, {Weiler}, {Wertz}, {Wevers}, {Wyrzykowski}, {Yoldas}, {{\v{Z}}erjal}, {Ziaeepour}, {Zorec}, {Zschocke}, {Zucker}, {Zurbach}, \& {Zwitter}}]{gaiacollab_2018}
{Gaia Collaboration}, {Brown}, A.~G.~A., {Vallenari}, A., {et~al.} 2018, \aap, 616, A1

\bibitem[{{H.~E.~S.~S. Collaboration} {et~al.}(2012){H.~E.~S.~S. Collaboration}, {Abramowski}, {Acero}, {Aharonian}, {Akhperjanian}, {Anton}, {Balenderan}, {Balzer}, {Barnacka}, {Becherini}, {Becker}, {Bernl{\"o}hr}, {Birsin}, {Biteau}, {Bochow}, {Boisson}, {Bolmont}, {Bordas}, {Brucker}, {Brun}, {Brun}, {Bulik}, {B{\"u}sching}, {Carrigan}, {Casanova}, {Cerruti}, {Chadwick}, {Charbonnier}, {Chaves}, {Cheesebrough}, {Cologna}, {Conrad}, {Couturier}, {Dalton}, {Daniel}, {Davids}, {Degrange}, {Deil}, {Dickinson}, {Djannati-Ata{\"\i}}, {Domainko}, {Drury}, {Dubus}, {Dutson}, {Dyks}, {Dyrda}, {Egberts}, {Eger}, {Espigat}, {Fallon}, {Farnier}, {Fegan}, {Feinstein}, {Fernandes}, {Fiasson}, {Fontaine}, {F{\"o}rster}, {F{\"u}{\ss}ling}, {Gajdus}, {Gallant}, {Garrigoux}, {Gast}, {G{\'e}rard}, {Giebels}, {Glicenstein}, {Gl{\"u}ck}, {G{\"o}ring}, {Grondin}, {H{\"a}ffner}, {Hague}, {Hahn}, {Hampf}, {Harris}, {Hauser}, {Heinz}, {Heinzelmann}, {Henri}, {Hermann}, {Hillert}, {Hinton}, {Hofmann}, {Hofverberg}, {Holler},
  {Horns}, {Jacholkowska}, {Jahn}, {Jamrozy}, {Jung}, {Kastendieck}, {Katarzy{\'n}ski}, {Katz}, {Kaufmann}, {Kh{\'e}lifi}, {Klochkov}, {Klu{\'z}niak}, {Kneiske}, {Komin}, {Kosack}, {Kossakowski}, {Krayzel}, {Laffon}, {Lamanna}, {Lenain}, {Lennarz}, {Lohse}, {Lopatin}, {Lu}, {Marandon}, {Marcowith}, {Masbou}, {Maurin}, {Maxted}, {Mayer}, {McComb}, {Medina}, {M{\'e}hault}, {Menzler}, {Moderski}, {Mohamed}, {Moulin}, {Naumann}, {Naumann-Godo}, {de Naurois}, {Nedbal}, {Nekrassov}, {Nguyen}, {Nicholas}, {Niemiec}, {Nolan}, {Ohm}, {de O{\~n}a Wilhelmi}, {Opitz}, {Ostrowski}, {Oya}, {Panter}, {Paz Arribas}, {Pekeur}, {Pelletier}, {Perez}, {Petrucci}, {Peyaud}, {Pita}, {P{\"u}hlhofer}, {Punch}, {Quirrenbach}, {Raue}, {Reimer}, {Reimer}, {Renaud}, {de los Reyes}, {Rieger}, {Ripken}, {Rob}, {Rosier-Lees}, {Rowell}, {Rudak}, {Rulten}, {Sahakian}, {Sanchez}, {Santangelo}, {Schlickeiser}, {Schulz}, {Schwanke}, {Schwarzburg}, {Schwemmer}, {Sheidaei}, {Skilton}, {Sol}, {Spengler}, {Stawarz}, {Steenkamp}, {Stegmann},
  {Stinzing}, {Stycz}, {Sushch}, {Szostek}, {Tavernet}, {Terrier}, {Tluczykont}, {Valerius}, {van Eldik}, {Vasileiadis}, {Venter}, {Viana}, {Vincent}, {V{\"o}lk}, {Volpe}, {Vorobiov}, {Vorster}, {Wagner}, {Ward}, {White}, {Wierzcholska}, {Zacharias}, {Zajczyk}, {Zdziarski}, {Zech}, \& {Zechlin}}]{hessj2012}
{H.~E.~S.~S. Collaboration}, {Abramowski}, A., {Acero}, F., {et~al.} 2012, \aap, 548, A46

\bibitem[{{H.~E.~S.~S. Collaboration} {et~al.}(2013){H.~E.~S.~S. Collaboration}, {Abramowski}, {Acero}, {Aharonian}, {Akhperjanian}, {Anton}, {Balenderan}, {Balzer}, {Barnacka}, {Becherini}, {Becker Tjus}, {Bernl{\"o}hr}, {Birsin}, {Biteau}, {Boisson}, {Bolmont}, {Bordas}, {Brucker}, {Brun}, {Brun}, {Bulik}, {Carrigan}, {Casanova}, {Cerruti}, {Chadwick}, {Chaves}, {Cheesebrough}, {Colafrancesco}, {Cologna}, {Conrad}, {Couturier}, {Dalton}, {Daniel}, {Davids}, {Degrange}, {Deil}, {deWilt}, {Dickinson}, {Djannati-Ata{\"\i}}, {Domainko}, {Drury}, {Dubus}, {Dutson}, {Dyks}, {Dyrda}, {Egberts}, {Eger}, {Espigat}, {Fallon}, {Farnier}, {Fegan}, {Feinstein}, {Fernandes}, {Fernandez}, {Fiasson}, {Fontaine}, {F{\"o}rster}, {F{\"u}{\ss}ling}, {Gajdus}, {Gallant}, {Garrigoux}, {Gast}, {Giebels}, {Glicenstein}, {Gl{\"u}ck}, {G{\"o}ring}, {Grondin}, {Grudzi{\'n}ska}, {H{\"a}er}, {Hague}, {Hahn}, {Hampf}, {Harris}, {Heinz}, {Heinzelmann}, {Henri}, {Hermann}, {Hillert}, {Hinton}, {Hofmann}, {Hofverberg}, {Holler}, {Horns},
  {Jacholkowska}, {Jahn}, {Jamrozy}, {Jung}, {Kastendieck}, {Katarzy{\'n}ski}, {Katz}, {Kaufmann}, {Kh{\'e}lifi}, {Klepser}, {Klochkov}, {Klu{\'z}niak}, {Kneiske}, {Kolitzus}, {Komin}, {Kosack}, {Kossakowski}, {Krayzel}, {Kr{\"u}ger}, {Lan}, {Lamanna}, {Lefaucheur}, {Lemoine-Goumard}, {Lenain}, {Lennarz}, {Lohse}, {Lopatin}, {Lu}, {Marandon}, {Marcowith}, {Masbou}, {Maurin}, {Maxted}, {Mayer}, {McComb}, {Medina}, {M{\'e}hault}, {Menzler}, {Moderski}, {Mohamed}, {Moulin}, {Naumann}, {Naumann-Godo}, {de Naurois}, {Nedbal}, {Nguyen}, {Niemiec}, {Nolan}, {Oakes}, {Ohm}, {de O{\~n}a Wilhelmi}, {Opitz}, {Ostrowski}, {Oya}, {Panter}, {Parsons}, {Paz Arribas}, {Pekeur}, {Pelletier}, {Perez}, {Petrucci}, {Peyaud}, {Pita}, {P{\"u}hlhofer}, {Punch}, {Quirrenbach}, {Raab}, {Raue}, {Reimer}, {Reimer}, {Renaud}, {de los Reyes}, {Rieger}, {Ripken}, {Rob}, {Rosier-Lees}, {Rowell}, {Rudak}, {Rulten}, {Sahakian}, {Sanchez}, {Santangelo}, {Schlickeiser}, {Schulz}, {Schwanke}, {Schwarzburg}, {Schwemmer}, {Sheidaei}, {Skilton},
  {Sol}, {Spengler}, {Stawarz}, {Steenkamp}, {Stegmann}, {Stinzing}, {Stycz}, {Sushch}, {Szostek}, {Tavernet}, {Terrier}, {Tluczykont}, {Trichard}, {Valerius}, {van Eldik}, {Vasileiadis}, {Venter}, {Viana}, {Vincent}, {V{\"o}lk}, {Volpe}, {Vorobiov}, {Vorster}, {Wagner}, {Ward}, {White}, {Wierzcholska}, {Willmann}, {Wouters}, {Zacharias}, {Zajczyk}, {Zdziarski}, {Zech}, \& {Zechlin}}]{psrb2011}
{H.~E.~S.~S. Collaboration}, {Abramowski}, A., {Acero}, F., {et~al.} 2013, \aap, 551, A94

\bibitem[{{H.E.S.S.\ Collaboration} {et~al.}(2020){H.E.S.S.\ Collaboration}, {Abdalla}, {Adam}, {Aharonian}, {Ait Benkhali}, {Ang{\"u}ner}, {Arakawa}, {Arcaro}, {Armand}, {Ashkar}, {Backes}, {Barbosa Martins}, {Barnard}, {Becherini}, {Berge}, {Bernl{\"o}hr}, {Blackwell}, {B{\"o}ttcher}, {Boisson}, {Bolmont}, {Bonnefoy}, {Bregeon}, {Breuhaus}, {Brun}, {Brun}, {Bryan}, {B{\"u}chele}, {Bulik}, {Bylund}, {Caroff}, {Carosi}, {Casanova}, {Cerruti}, {Chand}, {Chandra}, {Chaves}, {Chen}, {Colafrancesco}, {Cury{\l}o}, {Davids}, {Deil}, {Devin}, {deWilt}, {Dirson}, {Djannati-Ata{\"\i}}, {Dmytriiev}, {Donath}, {Doroshenko}, {Dyks}, {Egberts}, {Emery}, {Ernenwein}, {Eschbach}, {Feijen}, {Fegan}, {Fiasson}, {Fontaine}, {Funk}, {F{\"u}{\ss}ling}, {Gabici}, {Gallant}, {Gat{\'e}}, {Giavitto}, {Giunti}, {Glawion}, {Glicenstein}, {Gottschall}, {Grondin}, {Hahn}, {Haupt}, {Heinzelmann}, {Henri}, {Hermann}, {Hinton}, {Hofmann}, {Hoischen}, {Holch}, {Holler}, {Horns}, {Huber}, {Iwasaki}, {Jamrozy}, {Jankowsky}, {Jankowsky},
  {Jardin-Blicq}, {Jung-Richardt}, {Kastendieck}, {Katarzy{\'n}ski}, {Katsuragawa}, {Katz}, {Khangulyan}, {Kh{\'e}lifi}, {King}, {Klepser}, {Klu{\'z}niak}, {Komin}, {Kosack}, {Kostunin}, {Kreter}, {Lamanna}, {Lemi{\`e}re}, {Lemoine-Goumard}, {Lenain}, {Leser}, {Levy}, {Lohse}, {Lypova}, {Mackey}, {Majumdar}, {Malyshev}, {Malyshev}, {Marandon}, {Marcowith}, {Mares}, {Mariaud}, {Mart{\'\i}-Devesa}, {Marx}, {Maurin}, {Meintjes}, {Mitchell}, {Moderski}, {Mohamed}, {Mohrmann}, {Moore}, {Moulin}, {Muller}, {Murach}, {Nakashima}, {de Naurois}, {Ndiyavala}, {Niederwanger}, {Niemiec}, {Oakes}, {O'Brien}, {Odaka}, {Ohm}, {de Ona Wilhelmi}, {Ostrowski}, {Oya}, {Panter}, {Parsons}, {Perennes}, {Petrucci}, {Peyaud}, {Piel}, {Pita}, {Poireau}, {Priyana Noel}, {Prokhorov}, {Prokoph}, {P{\"u}hlhofer}, {Punch}, {Quirrenbach}, {Raab}, {Rauth}, {Reimer}, {Reimer}, {Remy}, {Renaud}, {Rieger}, {Rinchiuso}, {Romoli}, {Rowell}, {Rudak}, {Ruiz-Velasco}, {Sahakian}, {Sailer}, {Saito}, {Sanchez}, {Santangelo}, {Sasaki},
  {Schlickeiser}, {Sch{\"u}ssler}, {Schulz}, {Schutte}, {Schwanke}, {Schwemmer}, {Seglar-Arroyo}, {Senniappan}, {Seyffert}, {Shafi}, {Shiningayamwe}, {Simoni}, {Sinha}, {Sol}, {Specovius}, {Spir-Jacob}, {Stawarz}, {Steenkamp}, {Stegmann}, {Steppa}, {Takahashi}, {Tavernier}, {Taylor}, {Terrier}, {Tiziani}, {Tluczykont}, {Trichard}, {Tsirou}, {Tsuji}, {Tuffs}, {Uchiyama}, {van der Walt}, {van Eldik}, {van Rensburg}, {van Soelen}, {Vasileiadis}, {Veh}, {Venter}, {Vincent}, {Vink}, {V{\"o}lk}, {Vuillaume}, {Wadiasingh}, {Wagner}, {White}, {Wierzcholska}, {Yang}, {Yoneda}, {Zacharias}, {Zanin}, {Zdziarski}, {Zech}, {Zorn}, {{\.Z}ywucka}, \& {Bordas}}]{hess_psrb2020}
{H.E.S.S.\ Collaboration}, {Abdalla}, H., {Adam}, R., {et~al.} 2020, \aap, 633, A102

\bibitem[{{Huber} {et~al.}(2021){Huber}, {Kissmann}, {Reimer}, \& {Reimer}}]{2021A&A...646A..91H}
{Huber}, D., {Kissmann}, R., {Reimer}, A., \& {Reimer}, O. 2021, \aap, 646, A91

\bibitem[{{Johnson} {et~al.}(2018){Johnson}, {Wood}, {Kerr}, {Corbet}, {Cheung}, {Ray}, \& {Omodei}}]{Prsb2017Fermi}
{Johnson}, T.~J., {Wood}, K.~S., {Kerr}, M., {et~al.} 2018, \apj, 863, 27

\bibitem[{{Johnston} {et~al.}(2005){Johnston}, {Ball}, {Wang}, \& {Manchester}}]{johnston2005}
{Johnston}, S., {Ball}, L., {Wang}, N., \& {Manchester}, R.~N. 2005, \mnras, 358, 1069

\bibitem[{{Johnston} {et~al.}(1992){Johnston}, {Manchester}, {Lyne}, {Bailes}, {Kaspi}, {Qiao}, \& {D'Amico}}]{johnston92}
{Johnston}, S., {Manchester}, R.~N., {Lyne}, A.~G., {et~al.} 1992, \apjl, 387, L37

\bibitem[{{Johnston} {et~al.}(1996){Johnston}, {Manchester}, {Lyne}, {D'Amico}, {Bailes}, {Gaensler}, \& {Nicastro}}]{johnson96}
{Johnston}, S., {Manchester}, R.~N., {Lyne}, A.~G., {et~al.} 1996, \mnras, 279, 1026

\bibitem[{{Johnston} {et~al.}(1994){Johnston}, {Manchester}, {Lyne}, {Nicastro}, \& {Spyromilio}}]{johnston94}
{Johnston}, S., {Manchester}, R.~N., {Lyne}, A.~G., {Nicastro}, L., \& {Spyromilio}, J. 1994, \mnras, 268, 430

\bibitem[{{Kelly}(2007)}]{kelly_2007}
{Kelly}, B.~C. 2007, \apj, 665, 1489

\bibitem[{{Khangulyan} {et~al.}(2008{\natexlab{a}}){Khangulyan}, {Aharonian}, \& {Bosch-Ramon}}]{2008MNRAS.383..467K}
{Khangulyan}, D., {Aharonian}, F., \& {Bosch-Ramon}, V. 2008{\natexlab{a}}, \mnras, 383, 467

\bibitem[{Khangulyan {et~al.}(2012)Khangulyan, Aharonian, Bogovalov, \& Rib{\'{o} }}]{Khangulyan_2012}
Khangulyan, D., Aharonian, F.~A., Bogovalov, S.~V., \& Rib{\'{o} }, M. 2012, The Astrophysical Journal, 752, L17

\bibitem[{{Khangulyan} {et~al.}(2014){Khangulyan}, {Aharonian}, \& {Kelner}}]{2014ApJ...783..100K}
{Khangulyan}, D., {Aharonian}, F.~A., \& {Kelner}, S.~R. 2014, \apj, 783, 100

\bibitem[{Khangulyan {et~al.}(2007)Khangulyan, Hnatic, Aharonian, \& Bogovalov}]{khangulyan_2007}
Khangulyan, D., Hnatic, S., Aharonian, F., \& Bogovalov, S. 2007, Monthly Notices of the Royal Astronomical Society, 380, 320

\bibitem[{{Khangulyan} {et~al.}(2008{\natexlab{b}}){Khangulyan}, {Aharonian}, {Bogovalov}, {Koldoba}, \& {Ustyugova}}]{2008IJMPD..17.1909K}
{Khangulyan}, D.~V., {Aharonian}, F.~A., {Bogovalov}, S.~V., {Koldoba}, A.~V., \& {Ustyugova}, G.~V. 2008{\natexlab{b}}, International Journal of Modern Physics D, 17, 1909

\bibitem[{{Kirk} {et~al.}(1999){Kirk}, {Ball}, \& {Skj{\ae}raasen}}]{1999APh....10...31K}
{Kirk}, J.~G., {Ball}, L., \& {Skj{\ae}raasen}, O. 1999, Astroparticle Physics, 10, 31

\bibitem[{{Li} \& {Ma}(1983)}]{li_ma_1983}
{Li}, T.~P. \& {Ma}, Y.~Q. 1983, \apj, 272, 317

\bibitem[{{Marino} {et~al.}(2023){Marino}, {Driessen}, {Lenc}, {Rea}, {Malishev}, {Chernyakova}, {Kaplan}, {Murphy}, {Gaensler}, \& {Sivakoff}}]{marino_2023}
{Marino}, A., {Driessen}, L., {Lenc}, E., {et~al.} 2023, The Astronomer's Telegram, 15942, 1

\bibitem[{Miller-Jones {et~al.}(2018)Miller-Jones, Deller, Shannon, Dodson, Moldón, Ribó, Dubus, Johnston, Paredes, Ransom, \& Tomsick}]{millerjones_2018}
Miller-Jones, J. C.~A., Deller, A.~T., Shannon, R.~M., {et~al.} 2018, Monthly Notices of the Royal Astronomical Society, 479, 4849

\bibitem[{Negueruela {et~al.}(2011)Negueruela, Rib{\'{o} }, Herrero, Lorenzo, Khangulyan, \& Aharonian}]{Negueruela_2011}
Negueruela, I., Rib{\'{o} }, M., Herrero, A., {et~al.} 2011, The Astrophysical Journal, 732, L11

\bibitem[{Parsons \& Hinton(2014)}]{Parsons_2014}
Parsons, R. \& Hinton, J. 2014, Astroparticle Physics, 56, 26

\bibitem[{{Piron} {et~al.}(2001){Piron}, {Djannati-Atai}, {Punch}, {Tavernet}, {Barrau}, {Bazer-Bachi}, {Chounet}, {Debiais}, {Degrange}, {Dezalay}, {Espigat}, {Fabre}, {Fleury}, {Fontaine}, {Goret}, {Gouiffes}, {Khelifi}, {Malet}, {Masterson}, {Mohanty}, {Nuss}, {Renault}, {Rivoal}, {Rob}, \& {Vorobiov}}]{piron_2001}
{Piron}, F., {Djannati-Atai}, A., {Punch}, M., {et~al.} 2001, \aap, 374, 895

\bibitem[{{Shannon} {et~al.}(2014){Shannon}, {Johnston}, \& {Manchester}}]{shannon14}
{Shannon}, R.~M., {Johnston}, S., \& {Manchester}, R.~N. 2014, \mnras, 437, 3255

\bibitem[{{Sushch} \& {B{\"o}ttcher}(2014)}]{2014JHEAp...3...18S}
{Sushch}, I. \& {B{\"o}ttcher}, M. 2014, Journal of High Energy Astrophysics, 3, 18

\bibitem[{{Sushch} \& {van Soelen}(2017)}]{sushch17}
{Sushch}, I. \& {van Soelen}, B. 2017, \apj, 837, 175

\bibitem[{Sushch \& van Soelen(2023)}]{Sushch_2023}
Sushch, I. \& van Soelen, B. 2023, The Astrophysical Journal, 959, 30

\bibitem[{{Takata} {et~al.}(2012){Takata}, {Okazaki}, {Nagataki}, {Naito}, {Kawachi}, {Lee}, {Mori}, {Hayasaki}, {Yamaguchi}, \& {Owocki}}]{2012ApJ...750...70T}
{Takata}, J., {Okazaki}, A.~T., {Nagataki}, S., {et~al.} 2012, \apj, 750, 70

\bibitem[{{Tam} {et~al.}(2011){Tam}, {Huang}, {Takata}, {Hui}, {Kong}, \& {Cheng}}]{tam11}
{Tam}, P.~H.~T., {Huang}, R.~H.~H., {Takata}, J., {et~al.} 2011, \apjl, 736, L10

\bibitem[{{Tavani} \& {Arons}(1997)}]{tavani_1997}
{Tavani}, M. \& {Arons}, J. 1997, \apj, 477, 439

\bibitem[{{Tavani} {et~al.}(1994){Tavani}, {Arons}, \& {Kaspi}}]{1994ApJ...433L..37T}
{Tavani}, M., {Arons}, J., \& {Kaspi}, V.~M. 1994, \apjl, 433, L37

\bibitem[{{van Soelen} \& {Meintjes}(2015)}]{vansoelen_2015}
{van Soelen}, B. \& {Meintjes}, P.~J. 2015, \memsai, 86, 123

\bibitem[{{Weng} {et~al.}(2022){Weng}, {Qian}, {Wang}, {Torres}, {Papitto}, {Jiang}, {Xu}, {Li}, {Yan}, {Liu}, {Ge}, \& {Yuan}}]{weng22}
{Weng}, S.-S., {Qian}, L., {Wang}, B.-J., {et~al.} 2022, Nature Astronomy, 6, 698

\bibitem[{{Wood} {et~al.}(2018){Wood}, {Johnson}, {Ray}, {Kerr}, {Chernyakova}, \& {Fermi LAT Collaboration}}]{wood_2018}
{Wood}, K.~S., {Johnson}, T., {Ray}, P.~S., {et~al.} 2018, in American Astronomical Society Meeting Abstracts, Vol. 231, American Astronomical Society Meeting Abstracts \#231, 233.04

\bibitem[{Yoneda {et~al.}(2020)Yoneda, Makishima, Enoto, Khangulyan, Matsumoto, \& Takahashi}]{yoneda_2020}
Yoneda, H., Makishima, K., Enoto, T., {et~al.} 2020, Physical review letters, 125, 111103

\bibitem[{{Zabalza} {et~al.}(2013){Zabalza}, {Bosch-Ramon}, {Aharonian}, \& {Khangulyan}}]{2013A&A...551A..17Z}
{Zabalza}, V., {Bosch-Ramon}, V., {Aharonian}, F., \& {Khangulyan}, D. 2013, \aap, 551, A17

\end{thebibliography}

\afterpage{
\begin{appendices}

\section{The $\eta^{2}$ parameter}
\label{appendix:eta}
In order to to include the uncertainties of both X-ray and TeV data, we utilise a linear combination of $\chi^2$ tests ~\citep[see e.g.][]{bausch_2013}, denoted here as $\eta^2$, and defined as:
\begin{linenomath*}
\begin{equation}
    \eta^{2}(f,\Omega) = \sum\limits_i \frac{(X_{i}-f(T_i,\Omega))^{2}}{\delta X^{2}_{i}} + \sum\limits_i \frac{(T_{i}-f^{-1}(X_i,\Omega))^{2}}{\delta T^{2}_{i}}
\end{equation}
\end{linenomath*}
where $X_{i}$ and $T_{i}$ are the $i$-th X-ray and TeV flux values from the time-correlated dataset. Accordingly $\delta X_{i}$ and $\delta T_{i}$ are the uncertainties of these values. The dependency between X-ray and TeV fluxes was assumed to have the functional form $X = f(T,\Omega)$ with $\Omega$ standing for the variable parameter(s) of the function $f$. The inverse function $f^{-1}$ is given by: $T=f^{-1}(X,\Omega)$. For an accurate comparison between $\eta^2$ values, we also implement a method of reduced $\eta^2$, named $\Bar{\eta}^2$. After summing the reduction of the two constituent $\chi^2$ values, we reduce $\eta^2$ by applying
\begin{linenomath*}
\begin{align*}
\Bar{\eta}^2 = \frac{\eta^2}{2(N-1)}
\end{align*}
\end{linenomath*}
where $N$ is the number of correlated pairs.

We note that by design the $\eta^2$ test is symmetric with respect to the interchange of the $X\longleftrightarrow T$ datasets.
\end{appendices}
}

\end{document}